\documentclass[aps,prd,twocolumn,groupedaddress, showkeys, showpacs]{revtex4}
\usepackage{graphicx,subfigure,enumerate,pstricks,latexsym,longtable}
\usepackage[english]{babel}

\providecommand{\dif}{\mathrm{d}}
\def\beq{\begin{equation}}
\def\eeq{\end{equation}}
\def\bea{\begin{eqnarray}}
\def\eea{\end{eqnarray}}

\def\nn{\nonumber}

\newcommand{\Schw}{Schwarzschild}

\def\d{\dif}

\def\EE{{\cal E}}
\def\LL{{\cal L}}
\def\BB{{\cal B}}
\def\QQ{{\cal Q}}
\def\rr{r}
\def\tt{\theta}

\def\p{P}
\def\x{X}

\def\q{\tilde{q}}

\def\af{\zeta}
\def\dr{\delta r}
\def\dt{\delta \theta}

\def\RS{\Sigma}

\def\mir{\mathrm{r}}
\def\mit{\mathrm{\theta}}
\def\mip{\mathrm{\phi}}
\def\mil{\mathrm{L}}

\newcommand{\aaa}{{\diamond}}

\newcommand{\ce}{{\cal{E}}}
\newcommand{\cl}{{\cal{L}}}
\newcommand{\cb}{{\cal{B}}}

\begin{document}

\title{Circular orbits and related quasi-harmonic oscillatory motion of charged particles around weakly magnetized rotating black holes}
\author{Arman Tursunov}
\email{arman.tursunov@fpf.slu.cz}
\author{Zden\v{e}k Stuchl{\'i}k}
\email{zdenek.stuchlik@fpf.slu.cz}
\author{Martin Kolo\v{s}}
\email{martin.kolos@fpf.slu.cz}
\affiliation{Institute of Physics and Research Centre of Theoretical Physics and Astrophysics, Faculty of Philosophy and Science, Silesian University in Opava, \\
Bezru{\v c}ovo n{\'a}m.13, CZ-74601 Opava, Czech Republic}
\begin{abstract}
We study motion of charged particles in the field of a rotating black hole immersed into an external asymptotically uniform magnetic field, focusing on the epicyclic quasi-circular orbits near the equatorial plane. Separating the circular orbits into four qualitatively different classes according to the sign of the canonical angular momentum of the motion and the orientation of the Lorentz force, we analyse the circular orbits using the so called force formalism. We find the analytical solutions for the radial profiles of velocity, specific angular momentum and specific energy of the circular orbits in dependence on the black hole dimensionless spin and the magnetic field strength. The innermost stable circular orbits are determined for all four classes of the circular orbits. The stable circular orbits with outward oriented Lorentz force can extend to radii lower than the radius of the corresponding photon circular geodesic. We calculate the frequencies of the harmonic oscillatory motion of the charged particles in the radial and vertical directions related to the equatorial circular orbits and study the radial profiles of the radial, $\omega_\mir$, vertical, $\omega_\mit$, and orbital, $\omega_\mip$, frequencies, finding significant differences in comparison to the epicyclic geodesic circular motion. The most important new phenomenon is existence of toroidal charged particle epicyclic motion with $\omega_\mir \sim \omega_{\mit} >> \omega_{\mip}$ that could occur around retrograde circular orbits with outward oriented Lorentz force. We demonstrate that for the rapidly rotating black holes the role of the 'Wald induced charge' can be relevant.
\end{abstract}

\pacs{04.70.Bw, 04.25.-g, 04.70.-s, 97.60.Lf \hfill
}

\maketitle

\section{Introduction}

It is well known that magnetic fields have crucial role in accretion processes. In the Keplerian accretion discs \cite{Nov-Tho:1973:BlaHol:} local magnetic fields play a fundamental role in the viscosity mechanism of accretion due to the magneto-rotational instability \cite{Bal-Haw:1991:ApJ:}. In collisionless plasmas of accretion discs an internal global toroidal magnetic field could be created by the so called kinetic dynamo effect \cite{Cre-Stu:2013:PHYSRE:}. The kinetic phenomena could also govern transition from neutral to ionized equilibria of plasmas in accretion discs influenced by combined gravitational and electromagnetic fields \cite{Cre-Stu-Tes:2013:PhysPlasm:,Cre-Stu:2014:PhysPlasm:}. For example, in the kinetic approach we could model the equilibrium plasma configurations representing levitating tori \cite{Cre-etal:2013:ApJS:}. 

The black holes can be immersed in an external magnetic field that can have complex structure near the horizon, but at large distances can be approximated in a finite part of the space as close to a homogeneous magnetic field -- we use approximation of an asymptotically uniform magnetic field \cite{Kol-Stu-Tur:2015:CLAQG:}. For example, near the Galaxy centre containing the Sgr A* supermassive black hole a strong magnetic field has been detected \cite{Eat-etal:2013:NATUR:}. Such large-scale magnetic field could be generated during the early phases of the expansion of the Universe \cite{Rat:1992:ApJL:,Gra-Rub:2001:PhysRep:,Jai-Slo:2012:PRD:}. Further, a black hole near the equatorial plane of a magnetar can be immersed in a nearly uniform magnetic field, if the magnetar is at distance large enough \cite{Kov-etal:2014:PHYSR4:,Stu-Kol:2016:EPJC:}. 

The study of the charged test particle motion is considered to be the basis for understanding the influence of the magnetic fields on the accretion phenomena. For black holes carrying an electric charge and described by the Reissner-Nordstrom or Kerr-Newman geometry, the motion equations are separable and integrable \cite{Car:1973:BlaHol:}, giving a regular character of the motion that has been investigated in a number of papers \cite{Ruf:1973:Blahol:,Bic-Stu-Bal:1989:BAC:,Bal-Bic-Stu:1989:BAC:,Stu-Bic-Bal:1999:GRG:,Stu-Kot:2009:GRG:,Pug-Que-Ruf:2011:PHYSR4:,Pug-Que-Ruf:2013:PHYSR4:}. For weakly magnetized black holes, immersed in an external magnetic field represented by the Wald solution \cite{Wald:1974:PHYSR4:}, the equations of the motion are not separable and they have in general chaotic character. Various aspect of the motion of charged particles in the field of magnetized black holes were studied \cite{Prasanna:1980:RDNC:,Pre:2004:PHYSR4:,Fro-Sho:2010:PHYSR4:,Abd-Ahm-Jur:2013:PHYSR4:,Abd-Tur-Ahm-Kuv:2013:APSS:,Zah-etal:2013:PHYSR4:,Shi-Kim-Chi:2014:PHYSR4:,Kol-Stu-Tur:2015:CLAQG:,Sho:2015:ArXiv:}. Of special interest is existence of off-equatorial orbits \cite{Kov-Stu-Kar:2008:CLAQG:,Kov-etal:2010:CLAQG:,Kop-etal:2010:APJ:}, or the acceleration of particles of ionized Keplerian discs \cite{Stu-Kol:2016:EPJC:}. The 'magnetized' collisional processes describing acceleration of charged particles in the combined gravitational and electromagnetic fields \cite{Fro:2012:PHYSR4:,Iga-Har-Kim:2012:PHYSR4:,Stu-Sche-Abd:2014:PHYSR4:,Tos-Ahm-Abd-Stu:2014:PHYSR4:,Zas:2015:EPJC:} were shown to be able to reach large efficiency that could be obtained by uncharged particles in the superspinning geometry only \cite{Stu-Sche:2012:CLAQG:,Stu-Sche:2013:CLAQG:}. 

The purpose of the present paper is to study the motion of a charged test particle in vicinity of a weakly magnetized rotating (Kerr) black hole. For simplicity we assume the black hole immersed in an asymptotically uniform magnetic field with field lines parallel to the black-hole rotation axis. The uniform configuration of the magnetic field implies a simplified task, however, even in the axisymmetric background of such simply magnetized Kerr black hole the charged particle dynamics becomes non-integrable because of the absence of the Carter constant in the presence of the magnetic field, representing thus a complex problem. 

We demonstrate that presence of the external magnetic field generates four qualitatively different types of the circular orbits. We discuss properties of the circular orbits, giving especially the innermost stable circular orbits (ISCO) of the four types of the circular motion. In the previous works related to the motion of charged particles in the combined gravitational and magnetic fields, the problem had been solved by using the fully numerical methods or semi-analytical approaches \cite{Prasanna:1980:RDNC:,Pre:2004:PHYSR4:,Fro-Sho:2010:PHYSR4:,Abd-Ahm-Jur:2013:PHYSR4:}. In the present paper, we will apply a combination of the classical effective potential approach with the so called formalism of forces, giving thus an analytical form of the relevant equations. 

In the present paper we concentrate our attention mostly on the circular motion of charged particles and the related epicyclic motion. Assuming a slight deviation from purely circular character of the motion, we obtain frequencies of the radial and vertical harmonic or quasi-harmonic oscillatory motion of charged particles. Finally, we integrate the equations of the epicyclic motion and give the occurence of the new type of trajectories that could occur around magnetized Kerr black holes. 

This paper is organized as follows. In Sec.~\ref{WMKBH} we introduce the notion of the weakly magnetized Kerr black hole and discuss the limits of the applicability of the framework of the weak magnetization. The black hole rotation in the external magnetic field generates an induced charge which affects the motion of test particles. We show that the induced charge is weak in the sence that it does not modify the background Kerr spacetime. In Sec.~\ref{Sec-Dynamics} we study the dynamics of a charged test particle and separate the orbits into four qualitatively different classes. In Sec.~\ref{Sec-CO} we focus of the analysis of the charged particle circular orbits using the force formalism introduced in \cite{Abr-Nur-Wex:1995:CQG:,Kov-Stu:2007:CQG:}, and particularly we study the ISCO. In Sec.~\ref{Sec-Oscillations} we study the epicyclic motion. We summarize the results in Sec.~\ref{Summary}. 

Throughout the paper, we use the space-like signature (--,+,+,+) and the geometric system of units in which $G = 1 = c$. (However, for the expressions with an astrophysical application we use the units with the gravitational constant and the speed of light.) Greek indices are taken to run from 0 to 3; latin indices are related to the space components of the corresponding equations.

\section{Weakly magnetized Kerr black hole} \label{WMKBH}

We assume the magnetic field as an external and weak in such a sence that it cannot modify the metric of the background spacetime or, more precisely, the violation of the spacetime geometry by the magnetic field is negligibly small. Thus, we assume the geometry of the rotating black hole given by the Kerr metric 
\beq 
\d s^2 = g_{\mu\nu} \d x^{\mu} \d x^{\nu}, \label{KerrMetric}
\eeq
with the nonzero components of the metric tensor taking in the standard Boyer-Lindquist coordinates the form 
\bea 
g_{tt} = - \left( 1- \frac{2Mr}{\RS} \right), \quad
g_{t\phi} = - \frac{2Mra \sin^2\theta}{\RS} , \nonumber\\
g_{\phi\phi} = \left( r^2 +a^2 + \frac{2Mra^2}{\RS} \sin^2\theta \right) \sin^2\theta, \nonumber \\
g_{r r} = \frac{\RS}{\Delta}, \quad g_{\theta\theta} = \RS,
\label{MetricCoef} 
\eea
where 
\beq 
\RS = r^2 + a^2 \cos^2\theta, \quad \Delta = r^2 - 2Mr + a^2. 
\eeq
Here, $M$ is the gravitational mass of the black hole and $a$ is its spin parameter. The physical singularity is located at the ring with $r=0, \theta = \pi/2$ that can be well characterized in the so called Kerr-Schild "Cartesian" coordinates that are related to the Boyer-Lindquist coordinated by the relations  
\bea
x &=& (r^2+a^2)^{1/2}\sin \theta\cos\left[\phi-\tan^{-1}\left(\frac{a}{r}\right)\right],\\
y &=& (r^2+a^2)^{1/2}\sin \theta\sin\left[\phi-\tan^{-1}\left(\frac{a}{r}\right)\right],\\
z &=& r\cos\theta .
\eea
At the $x$--$z$ plane, the physical singularity is located at $x=\pm a$ and $z=0$.

In the following, we consider only the external regions of the Kerr black hole spacetimes located above the outer horizon ($r > r_{+}$, $a^2 < M^2$) where the ring singularity and the causality violations region of the Kerr spacetime are irrelevant. The outer horizon is located at 
\beq
r_{+} = M + (M^2-a^2)^{1/2}.
\eeq
The static limit surface $r_{\rm stat}(\theta)$, governing the boundary of the ergosphere, is given by 
\beq
r_{\rm stat}(\theta) = M + (M^2-a^2 \cos^2\theta)^{1/2} .
\eeq
The most convenient systems for treating the physical processes around rotating black holes are the so called locally non-rotating frames (LNRF) that correspond to the zero-angular-momentum observers (ZAMO) \cite{Bar-Pre-Teu:1972:ApJ:}. The 4-velocity of ZAMO is given by the formula 
\beq
 n^{\alpha} = (n^t,0,0,n^\phi),  \label{uZAMO}
\eeq
where
\beq
 (n^t)^2=\frac{g_{\phi\phi}}{g_{t\phi}^2-g_{tt}g_{\phi\phi}}, \quad n^\phi=-\frac{g_{t\phi}}{g_{\phi\phi}}\,n^t.
\eeq

The assumption of weakness of the external magnetic field can be applied, if the strength of the magnetic field satisfies the condition \cite{Fro-Sho:2010:PHYSR4:} 
\beq
B << B_{\rm G} =  {c^4\over G^{3/2} M_{\odot}}\left(\frac{M_{\odot}}{M}\right)\sim 10^{19}{M_{\odot}\over M}\mbox{Gauss}\, .
\label{BBB}
\eeq
The value of $B_{\rm G}$ in the estimation (\ref{BBB}) comes from the comparison of the gravitational effect of a black hole mass $M$ with the effect of the magnetic field $B$ on the spacetime curvature. For most of the astrophysical black holes the condition (\ref{BBB}) is perfectly satisfied. For instance, in the magnetic coupling processes studied in \cite{Pio-etal:2011:ASBULL:}, based on the use of the fundamental variability plane, the estimations of the magnitude of the magnetic field in the black hole vicinity give the values 
\bea
&& B\approx 10^8 \mbox{Gauss, \quad  for} \quad  M\approx 10 M_{\odot},\\
&& B\approx 10^4 \mbox{Gauss, \quad for} \quad M\approx 10^9 M_{\odot},
\eea
that are many orders of magnitude less than the value of $B_{\rm G}$. However, the effect of magnetic field which is weak in comparison to the gravitational mass effect (\ref{BBB}) in the case of the spacetime curvature, can be quite large for the motion of charged test particles. The relative influence of the magnetic field induced by the Lorentz force $q B/(m c)$ is governed by the specific charge of the particle (ratio of the electric charge and mass of the particle) and is of the order of 
\beq
b \sim 4.7 \times 10^7 \left( \frac{q}{e}\right) \left(\frac{m}{m_{\rm p}}\right)^{-1} \left( \frac{B}{10^8 {\rm G}}\right) \left(\frac{M}{10 M_{\odot}} \right), \label{LorB}
\eeq
where $m_{\rm p}$ is the proton mass. Thus, for the astrophysically relevant black holes the expression (\ref{LorB}) is large and cannot be neglected. 
 
Due to the stationarity and axial symmetry of the Kerr black hole spacetime, the vector potential of the weak magnetic field considered in \cite{Wald:1974:PHYSR4:}, which is solution of the vacuum Maxwell equations with Lorentz calibrated potential, $A^{\mu}_{;\mu} = 0$, can be chosen as linear combination of the spacetime Killing vectors 
\beq 
A^{\alpha} = C_1 \xi_{(t)}^{\alpha} + C_2 \xi_{(\phi)}^{\alpha}, \label{VecPot}
\eeq
where $\xi_{(t)} = \partial/\partial t$ and $\xi_{(\phi)} = \partial/\partial \phi$ are the timelike and spacelike axial Killing vectors which are reflecting the stationarity and axial symmetry of the background metric (\ref{KerrMetric}). Since the magnetic field is weak and can be described as a test field, we can freely choose the configuration of the magnetic field. According to \cite{Wald:1974:PHYSR4:}, we can specify the constants $C_1$ and $C_2$ of (\ref{VecPot}) as 
\beq 
C_1 = a B, \quad C_2 = \frac{B}{2}, 
\eeq
for the asymptotically uniform magnetic field with the strength $B$ directed along the axis of symmetry of the spacetime. The parameters $C_1$ and $C_2$ can be easily obtained from the asymptotic properties and the conditions of the electrical neutrality of the source and the uniformity of the external magnetic field. Thus, the non-zero components of the four-vector potential of the asymptotically uniform magnetic field take the form 
\beq 
A_t = \frac{B}{2} \left(g_{t\phi} + 2 a g_{tt}\right), \quad A_{\phi} =  \frac{B}{2} \left(g_{\phi\phi} + 2 a g_{t\phi}\right).
\label{VecPotTP}
\eeq
The terms proportional to the rotation parameter $a$ give the contribution to the Faraday induction which generates the electric potential and thus produces an induced electric field \cite{Wald:1974:PHYSR4:}. The potential difference between the horizon of a black hole and infinity takes the form 
\beq
\Delta \varphi = \varphi_{\rm H} - \varphi_{\infty} = \frac{Q - 2 a M B}{2 M}.
\eeq
This causes selective accretion of charged particles into the rotating black hole. The process is similar to those of the field generated by the rotating conductor immersed in a magnetic field. At the stage of the selective accretion which neutralizes the black hole, the four-vector potential of the resulting electromagnetic field takes the form 
\beq 
A^{\alpha} = \frac{B}{2} \left(\xi_{(\phi)}^{\alpha} + 2 a \xi_{(t)}^{\alpha}\right) - \frac{Q}{2M} \xi_{(t)}^{\alpha}. \label{VecPotAcr}
\eeq

Thus, the expressions (\ref{VecPotTP}) for the non-zero covariant components of the four-vector potential should be rewritten as \cite{Fro-Sho:2010:PHYSR4:,Kov-etal:2010:CLAQG:} 
\bea 
A_t &=& \frac{B}{2} \left(g_{t\phi} + 2 a g_{tt}\right) - \frac{Q}{2 M} g_{tt} \label{VecPotTPQ1} \\
A_{\phi} &=&  \frac{B}{2} \left(g_{\phi\phi} + 2 a g_{t\phi}\right) - \frac{Q}{2 M} g_{t\phi}. \label{VecPotTPQ2}
\eea 
The process of selective accretion occurs very fastly for the astrophysical black holes until the potential difference vanishes which means that the black hole obtains an inductive charge $Q_{\rm W} = 2 a M B$. The charge $Q_{\rm W}$ for the parallel orientation of the spin of a black hole $a$ and the magnetic field $B$ had been introduced by Wald \cite{Wald:1974:PHYSR4:}. Substituting the Wald charge into Eq.(\ref{VecPotAcr}) one obtains the expression for the four-vector potential after the process of selective accretion completed
\beq   
A^{\alpha} = \frac{B}{2} \xi_{(\phi)}^{\alpha}. \label{VecPot3} 
\eeq

Hereafter in the paper we will use the most general form of the four vector potential (\ref{VecPotAcr}). However, in particular cases we will specify the charge $Q$, considering two limit scenarios: 
\begin{itemize}
\item Black hole with $Q = Q_0 = 0$,
\item Black hole with Wald charge $Q=Q_{\rm W}=2 a M B$.
\end{itemize}  

One can compare the characteristic length scale given by the charge of the Reissner-Nordstrom black hole $Q_{\rm G}$ with its gravitational radius
\beq 
\sqrt{\frac{Q_{\rm G}^2 G}{c^4}} = \frac{2 G M}{c^2}. 
\eeq
This gives the charge, whose gravitational effect is comparable with the spacetime curvature of a black hole. For the black hole of mass $M$ this condition implies that the gravitational effect of the charge $Q$ on the background geometry can be neglected if 
\beq 
Q << Q_{\rm G} = 2 G^{1/2} M \approx 10^{30} {M\over M_{\odot}} ~\mbox{statC}. \label{QG} 
\eeq 
The value of the Wald charge $Q_{\rm W} = 2 M a B \leq 2 M^2 B$ is
\beq 
Q_{\rm W} \leq 10^{18} \left({M\over M_{\odot}}\right)^2 \left(\frac{B}{10^8 \mbox{G}}\right) ~\mbox{statC}, 
\eeq
which obviously satisfies the condition (\ref{QG}). This implies that the induced charge of the rotating black hole is weak in the same sense as the external magnetic field, namely, it cannot modify the background geometry of the black hole. 
  
Hereafter in this paper we will use for simplicity the system of units in which the mass of the black hole is equal to unity, $M=1$, i.e., we express the related quantities in units of the black hole mass. 


\section{Dynamics of charged particles} \label{Sec-Dynamics}

\subsection{Equations of motion} \label {Sec-EOM}

In this section we consider motion of a charged particle of mass $m$ and electric charge $q$ in the field of an axially symmetric rotating (Kerr) black hole immersed in an external asymptotically uniform magnetic field with field lines oriented in the direction of the black hole rotation axis. The motion of charged particles is governed by the Lorentz equation. Due to the assumption of the symmetries of the combined gravitational and electromagnetic background of the magnetized black hole, we can efficiently use the Hamiltonian formalism. Such an assumption allows for substantial simplification of the equations of motion enabling to find simple solutions of the charged particle motion that give an insight into the physical phenomena occuring in the combined gravitational and electromagnetic fields. The dynamical equations for the neutral particle motion can be obtained by taking vanishing charge of the particle, $q=0$. 

The Hamiltonian for dynamics of a charged particle can be written in the form
\beq
  H =  \frac{1}{2} g^{\alpha\beta} (\p_\alpha - q A_\alpha)(\p_\beta - q A_\beta) + \frac{1}{2} \, m^2
  \label{particleHAM},
\eeq
where the kinematical four-momentum $p^{\mu} = m u^{\mu}$ is related to the generalized (canonical) four-momentum $P^{\mu}$ by the relation
\beq
 P^\mu = p^{\mu} + q A^{\mu}. \label{particleMOM}
\eeq
The dynamics of charged particles governed by the Hamiltonian (\ref{particleHAM}) is given by the Hamilton equations 
\beq
 \frac{\d \x^\mu}{\d \af} = \frac{\partial H}{\partial P_\mu}, \quad
 \frac{\d P_\mu}{\d \af} = - \frac{\partial H}{\partial \x^\mu}, \label{Ham_eq}
\eeq 
where we introduced affine parameter $\af$ related to particle proper time $\tau$ by the relation $\af=\tau/m$.

Using the symmetries of the background spacetime (\ref{KerrMetric}) and the uniformity of the asymptotic configuration of the magnetic field, one can easily find the existing conserved quantities related to the charged particle which are the specific energy ${\cal E}$ and specific angular momentum ${\cal L}$ that are given in terms of the metric coefficients (\ref{MetricCoef}) and the vector potential (\ref{VecPotAcr}): 
\bea
- {\cal E} \equiv - \frac{E}{m} = \xi^{\mu}_{(t)} \frac{P_{\mu}}{m} = g_{tt} \frac{dt}{d\tau} + g_{t\phi}\frac{d\phi}{d\tau} + \frac{q}{m} A_{t},&& \label{Energy} \\
{\cal L} \equiv \frac{L}{m} = \xi^{\mu}_{(\phi)} \frac{P_{\mu}}{m} = g_{\phi\phi} \frac{d\phi}{d\tau} + g_{t\phi}\frac{dt}{d\tau} + \frac{q}{m} A_{\phi}.&& \label{AngMom} 
\eea

Using constant of the motion $\cl$ and $\ce$, and specific charge $\tilde{q}=q/m$, we can re-write the Hamiltonian (\ref{particleHAM}) as 
\beq 
H = \frac{1}{2} g^{rr} \, p_r^2 + \frac{1}{2} g^{\theta\theta} \, p_\theta^2 + H_{\rm P}(r,\theta),
\eeq
where the potential part of the Hamiltonian $H_{\rm P}(r,\theta)$ is introduced in the form 
\bea
 H_{\rm P} &=& \frac{1}{2}\big[ g^{tt} (\ce + \tilde{q} A_{t})^2 - 2 g^{t\phi}(\ce + \tilde{q} A_{t}) (\cl-\tilde{q} A_{\phi}) \nn \\
 && + g^{\phi\phi} (\cl-\tilde{q} A_{\phi})^2 + 1 \big] \label{HamPot}.
\eea

From the equations (\ref{Energy}) and (\ref{AngMom}) we obtain the first two equations of motion of charged particle in the form 
\bea
\frac{dt}{d\tau} &=& - \frac{g_{\phi\phi} (\EE+\q A_t) + g_{t\phi} (\LL-\q A_{\phi})}{g_{tt} g_{\phi\phi} - g_{t\phi}^2}, \label{tequat} \\
\frac{d\phi}{d\tau} &=& \frac{g_{tt} (\LL-\q A_{\phi}) + g_{t\phi} (\EE+\q A_t)}{g_{tt} g_{\phi\phi} - g_{t\phi}^2} . 
\label{phiequat} 
\eea
In order to find the remained equations of the motion analytically, we will concentrate our study to the motion in the equatorial plane $\theta = \pi/2$, $\dot{\theta} = 0$ and use the normalization condition $u^{\alpha} u_{\alpha} = -1$ (equivalent to $H=0$). Then the equation of the radial motion of the charged particle in the combined gravitational and magnetic fields will take the form
\beq  
\left(\frac{dr}{d\tau}\right)^2  = \frac{R (r)}{r^3}, \label{radialmotion}
\eeq
where for a simple representative form we define the radial function $R(r)$ governing the radial motion of charged test particles in terms of the components of the metric tensor (\ref{MetricCoef}) as follows 
\beq 
R (r;a,B,\EE,\LL) = - 2 r^3 g^{rr}(r;a) \, H_{\rm P}(r;a,B,\EE,\LL). \label{RfuncDef} 
\eeq

\begin{figure}
\includegraphics[width=1.\hsize]{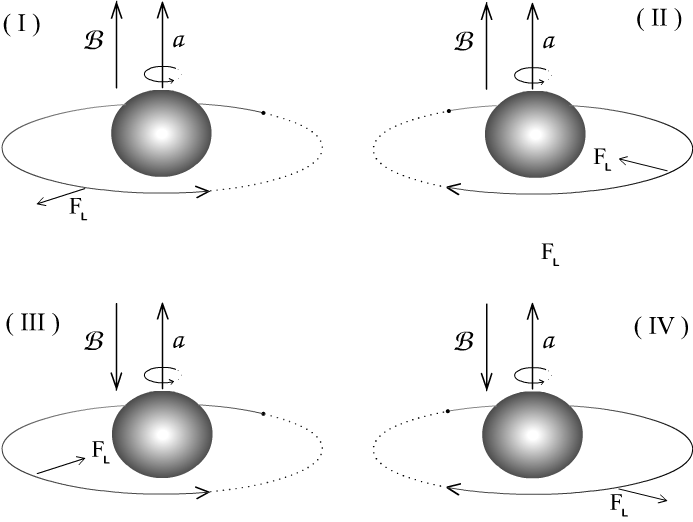}
\caption{ \label{Schema}
Representation of the four classes of the circular motion of charged particles. ${\rm F_{\rm L}}$ indicates orientation of the Lorentz force. 
}
\end{figure}

Let us analyze the symmetry features of the radial function governing the radial motion (\ref{RfuncDef}). The equations (\ref{tequat}) - (\ref{RfuncDef}) are invariant under the transformations $a \rightarrow - a$, $\q B \rightarrow - \q B$, $L \rightarrow - L$ and the re-definition of the axial coordinate $\phi \rightarrow - \phi$. Instead of using $\q B$ it is useful to use the following magnetic field parameter \cite{Fro:2012:PHYSR4:} and charge parameter
\beq
\BB=\frac{qB}{2m}, \quad \QQ=\frac{qQ}{m}. 
\eeq 
Hereafter we consider the black hole in two special cases, namely with zero induced charge $\QQ=\QQ_0=0$ and with 'Wald induced charge' $\QQ=\QQ_{\rm W}=4a\BB$.

Without loss of generality, we can take the specific charge of the particle $\q$ and the parameter of the rotation of a black hole $a$ as always positive. For a particle with negative charge it is sufficient to use the transformations given above. 

\subsection{Four types of equatorial circular orbits} \label{Sec-Types}

The circular motion of charged particles occurs in prograde (or co-rotating) orbits with the canonical angular momentum $\LL>0$, and retrograde (or counter-rotating) orbits with $\LL<0$. On the other hand, for each case the presence of an external magnetic field produces the Larmor and anti-Larmor orbits corresponding to the Lorentz force acting on the charged particles toward the black hole or in the outward direction, respectively \cite{Ali-Ozd:2002:MNRAS:}. We accept the rotation parameter of a black hole as always positive $a \geq 0$ as well as the charge of the test particle $q>0$. This implies that we can distinguish four different types of circular motion for the charged particle in the magnetized Kerr black hole spacetime, in contrast to the magnetized Schwarzschild black hole case where we can have only two different configurations. The four types of the charged particle motion in the equatorial plane of the magnetized Kerr black holes, represented in Fig.\ref{Schema}, are given in the following way  
\begin{itemize}
\item[(I)] {\it Prograde anti-Larmor orbits (PALO)} corresponding to $\cl>0, \cb>0$. Magnetic field lines are oriented in the same direction as the rotation axis of the black hole. The Lorentz force acting on a charged particle {\it co-rotating} with the black hole is {\it repulsive}, i.e., directed outwards the black hole. 
\item[(II)] {\it Retrograde Larmor orbits (RLO)} corresponding to $\cl<0, \cb>0$. Magnetic field lines are oriented in the same direction as the rotation axis of the black hole. The Lorentz force acting on a {\it counter-rotating} charged particle is {\it attractive}, i.e., directed towards the black hole. 
\item[(III)] {\it Prograde Larmor orbits (PLO)} corresponding to $\cl>0, \cb<0$. Magnetic field lines are oriented in opposite direction with respect to the rotation axis of the black hole. The Lorentz force acting on a {\it co-rotating} charged particle is {\it attractive}.
\item[(IV)] {\it Retrograde anti-Larmor orbits (RALO)} corresponding to $\cl<0, \cb<0$. Magnetic field lines are oriented in opposite direction with respect to the rotation axis of the black hole. The Lorentz force acting on a  {\it counter-rotating} charged particle is {\it repulsive}.
\end{itemize}
Note, that the signs in the definition of the types of orbits are valid for the positive values of the rotational parameter, $a \geq 0$. For negative values of $a$ it is sufficient to make the tranformations discussed below Eq.(\ref{RfuncDef}). 

The circular orbits play an important role in understanding the essential features of the dynamics of test particles around a rotating black hole immersed in an uniform magnetic field. Physically, from the symmetry of the problem, it is clear that circular orbits are possible in the equatorial plane where $\theta = \pi/2$; and they further require $dr/d\tau = 0$. The existence of the circular orbits requires vanishing of the radial function $R$ given by (\ref{RfuncDef}), along with its first derivative with respect to the radial coordinate $r$ 
\beq 
     R = 0, \qquad \partial_r R = 0. \label{cir-cond} 
\eeq

The direct solution of these equations would determine the energy and the axial angular momentum of a charged particle at the circular orbit in terms of the orbital radius $r$, the black hole dimensionless spin $a$, and the magnetic field parameter $\BB$. However, the expressions (\ref{cir-cond}) are high-order polynomial equations and the analytical solution of (\ref{cir-cond}) is very complex and cannot be presented in a reasonable representative form. Therefore, in some papers, e.g. \cite{Prasanna:1980:RDNC:}, a numerical analysis of the above presented equations is performed. The study of the charged particle motion around a Kerr black hole immersed in an uniform magnetic field realized through the analysis of the radial function of the radial motion is not effective. For this reason, in the next section related to the study of the circular orbits we will combine the above presented standard approach with the so called force formalism suggested in \cite{Abr-Nur-Wex:1995:CQG:} and well applied in \cite{Kov-Stu-Kar:2008:CLAQG:,Kov-etal:2010:CLAQG:,Kov-Stu:2007:CQG:} for the study of the particle motion in the field of magnetized Kerr black holes. 


\section{ Circular motion of charged particles } \label{Sec-CO}


\subsection{Formalism of forces}

\begin{figure*}
\includegraphics[width=\hsize]{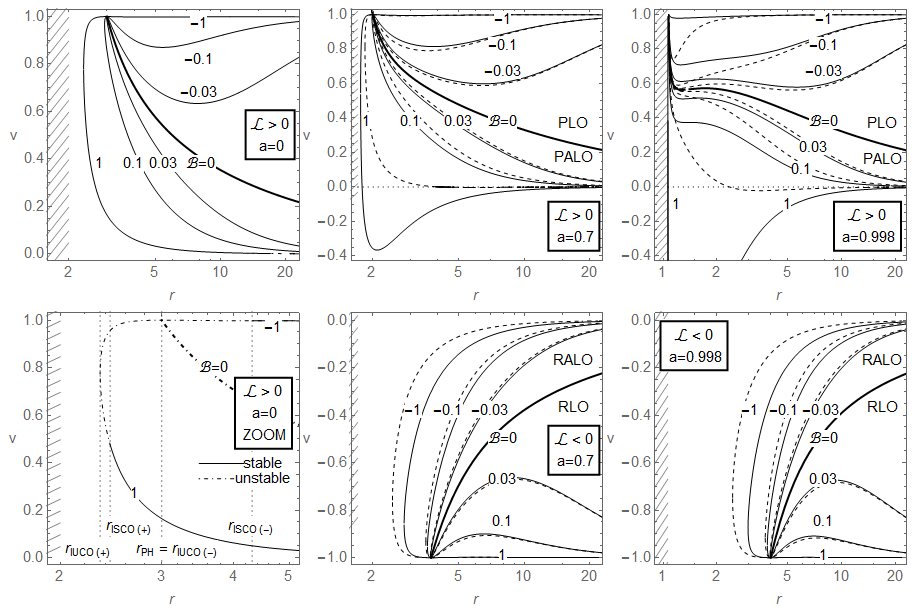}
\caption{\label{Veloc1} Radial profiles of the LNRF velocity of the charged particle at the circular orbits around magnetized black holes given for representative values of the spin $a$ and the magnetic field parameter $\BB$. The left column of plots corresponds to the non-rotating black holes with $a=0$, the middle column of plots represent prograde and retrograde orbits of rotating black holes with $a=0.7$ while the right column with $a=0.998$. The thick solid curves correspond to the non-magnetized black holes with $\BB=0$, separating the regions with the Larmor and anti-Larmor motions. The solid curves correspond to the case with Wald charge $\QQ=\QQ_{\rm W}$, while the dashed curves represent the case with zero charge $\QQ=0$. The dotted line give the position of photon orbit and the positions for innermost unstable circular orbit ($r_{\rm IUCO}$ and innermost stable circular orbit $r_{\rm ISCO}$ in the case of non-rotating black hole.
}
\end{figure*}
\begin{figure*}
\includegraphics[width=\hsize]{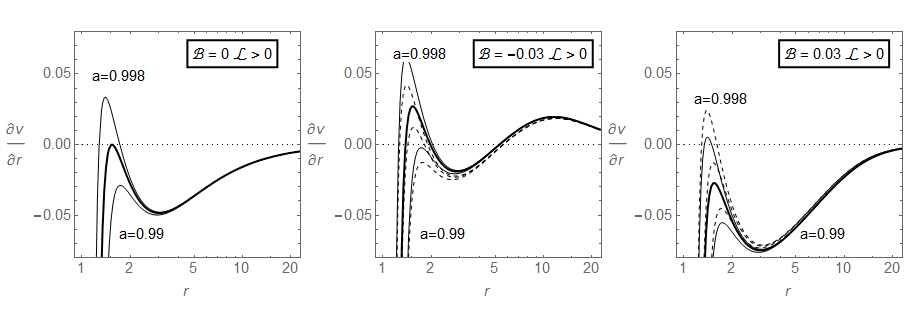}
\caption{ \label{Veloc-grad} Radial profiles of the gradient of the LNRF velocity. Since the Aschenbach effect is observed in the prograde Larmor motion only, we plotted PALO and PLO cases only. The left figure represents the pure Kerr spacetimes where thin solid curves represent black hole spins $a=0.99$ and $a=0.998$, while the thick solid curve represents the Aschenbach effect limit for $a=0.9953$. The middle figure represents the prograde Larmor orbits, while the right figure represents the prograde anti-Larmor orbits. Even small value of the magnetic field parameter, $\BB=\pm0.03$, can remarkably change the limiting black hole spin of the Aschenbach effect in the field of magnetized Kerr black holes. 
}
\end{figure*}

We can describe the charged particle motion directly by the Lorentz equation
\beq
u^{\mu}\nabla_{\mu} u^{\nu}=\q\,F^{\nu}_{\mu}u^{\mu}, \label{Lorentz}
\eeq
where $\q$ is the specific charge, $u^{\mu}$ is the four-velocity of the particle normalized by the condition $u^{\mu} u_{\mu} = -1$ and $F_{\mu \nu} = A_{\nu,\mu} - A_{\mu,\nu}$ is the antisymmetric tensor of the considered electromagnetic field. To achieve the purposes of the present paper, we use the formalism of forces \cite{Abr-Nur-Wex:1995:CQG:}, which is based 
on the projection  of the Lorentz equation (\ref{Lorentz}) onto the three-dimensional hypersurface $h_{ik} = g_{ik} + n_{i} n_{k}$, orthogonal to the four-velocity field of the locally non-rotating frames (LNRF) \cite{Abr-Nur-Wex:1995:CQG:,Stu-Kov:2006:IJMPD:} 
\bea
&& n^{\mu}=e^{-\Phi}(\xi_{(t)}^{\mu}+\Omega_{_{\rm LNRF}} \xi_{(\phi)}^{\mu} ), \\
&& e^{2\Phi}=-(\xi_{(t)}^{\mu}+\Omega_{_{\rm LNRF}}\xi_{(\phi)}^{\mu})(\xi_{(t)}^{\mu}+\Omega_{_{\rm LNRF}} \xi_{(\phi)}^{\mu}),
\eea
where the angular velocity of the LNRF is 
\beq
\Omega_{_{\rm LNRF}}=-g_{t\phi}/g_{\phi\phi}. 
\eeq
Vectors $\xi_{(t)}^{\mu}$ and $\xi_{(\phi)}^{\mu}$ correspond to the timelike and spacelike Killing vecors defined after Eq.(\ref{VecPot}). In the spherically symmetric spacetimes, $\Omega_{_{\rm LNRF}}=0$. 

The four-velocity field of the charged test particles uniformly revolving along the circular orbits can be written as 
\beq
u^{\mu}=\gamma(n^{\mu}+v \tau^{\mu}), \label{A1}
\eeq
where, $\gamma=(1-v^2)^{-1/2}$ is the Lorentz gamma factor, $\tau^{\mu}=\xi_{(\phi)}^{\mu} \,g_{\phi\phi}^{-1/2}$ is a unit spacelike vector orthogonal to $n^{\mu}$ (here considered to be the 4-velocity of the ZAMO, i.e., LNRF given by Eq.~(\ref{uZAMO})), along which the spatial velocity $v^{\mu}=v\tau^{\mu}$ is aligned. In general, the vectors in the expression (\ref{A1}) correspond to the standard orthonormal tetrad applied to the LNRF as $n^{\mu}=e_{(t)}^{\mu}$ and $\tau^{\mu}=e_{(\phi)}^{\mu}$. Thus, $v$ is the orbital or azimuthal velocity measured with respect to the LNRF. The LNRF components of the electromagnetic field tensor related to the asymptotically uniform magnetic field are given in the Appendix. 

Projection of the Lorentz equation, $h^k_j u^i\nabla_i u_k=\q h^i_j F_{ik}u^k$, can be written in the form \cite{Kov-etal:2010:CLAQG:} 
\beq
\label{Force-equation}
\mathcal{G}_a+(\gamma v)^2\mathcal{Z}_a+\gamma^2v\mathcal{C}_a=-\gamma(\mathcal{E}_a+v\mathcal{M}_a),
\eeq
where from the left hand side the so-called mass and velocity independent parts of the gravitational $\mathcal{G}$, centrifugal $\mathcal{Z}$ and Coriolis $\mathcal{C}$ inertial forces, and from the right hand side, the electric $\mathcal{E}$ and magnetic $\mathcal{M}$ forces can be expressed as 
\bea
\mathcal{G}_a &=& -\partial_a\Phi,\\
\mathcal{Z}_a &=& \textstyle{\frac{1}{2}}g_{\phi\phi}^{-1}\,e^{-2\Phi}\Big(e^{2\Phi}\,\partial_{a}g_{\phi\phi}-g_{\phi\phi}\,\partial_a e^{2\Phi}\Big),\\
\mathcal{C}_a &=& g_{\phi\phi}^{-3/2}\,e^{-\Phi}\Big(g_{\phi\phi}\,\partial_{a}g_{t\phi}-g_{t\phi}\,\partial_{a} g_{\phi\phi}\Big),\\
\mathcal{E}_a &=& \q~ e^{-\Phi}\Big(\Omega_{_{\rm LNRF}}\partial_{a} A_{\phi}+\partial_{a} A_t\Big),\\
\mathcal{M}_a &=& \q~ g_{\phi\phi}^{-1/2}\,\partial_{a} A_{\phi},
\eea
where only the radial $r$ and latitudinal $\theta$ components are nonzero, thus the index $a = r, \theta$. In the case of a Kerr black hole immersed in an asymptotically uniform magnetic field, the radial components of the gravitational $\mathcal{G}_r$, centrifugal $\mathcal{Z}_r$, Coriolis $\mathcal{C}_r$, electric $\mathcal{E}_r$ and magnetic $\mathcal{M}_r$ forces take the following form
\bea
\mathcal{G}_r&=&-\frac{a^4+2 a^2 (r-2) r+r^4}{\mu^2 \nu^2},\\
\mathcal{Z}_r&=&\frac{r \left(a^2+3 r^2\right)}{\mu^2}+\frac{1-r}{\nu^2}-\frac{1}{r},\\
\mathcal{C}_r&=&\frac{2 a \left(a^2+3 r^2\right)}{\mu^2 \nu },\\
\mathcal{E}_r&=&\frac{2 a \BB \left(r^2-a^2\right) + \QQ \left(r^2+a^2\right)}{\mu  \nu r },\\
\mathcal{M}_r&=&\frac{ 2 \BB \left(r^3 + a^2\right) + \QQ \left(r^2+a^2\right)}{\mu  r},
\eea
where
\beq 
\mu =\sqrt{r \left(a^2 (r+2)+r^3\right)}, \quad \nu = \sqrt{a^2+(r-2) r}. 
\eeq
Remind that in the equations given above and hereafter we fix and assume the mass of the black hole to be equal to unity, $M=1$. Fixing the plane of the motion to the equatorial plane, we can determine, according to the force formalism, the axial angular momentum of a charged particle at the circular orbit, $\LL$, from the radial component of the equation (\ref{Force-equation}) 
\beq
\mathcal{G}_r+(\gamma v)^2\mathcal{Z}_r+\gamma^2 v \mathcal{C}_r=-\gamma(\mathcal{E}_r+v\mathcal{M}_r). \label{ForcesBalance}
\eeq
Thus we get the fourth order equation in the LNRF orbital velocity $v$ 
\beq
Av^4+Cv^3+Dv^2+F v + H=0, \label{polinom}
\eeq
where
\bea
A&=&(\mathcal{G}_r-\mathcal{Z}_r)^2+\mathcal{M}_{r}^2,\\
C&=& 2 \mathcal{C}_{r}(\mathcal{Z}_{r}-\mathcal{G}_{r})+2 \mathcal{E}_{r} \mathcal{M}_{r},\\
D&=&\mathcal{C}_{r}^2 + 2\mathcal{G}_{r} (\mathcal{Z}_{r}-\mathcal{G}_{r}) + (\mathcal{E}_{r}^2 - \mathcal{M}_{r}^2),\\
F&=& 2 (\mathcal{C}_{r}\mathcal{G}_{r} - \mathcal{E}_{r}\mathcal{M}_{r}),\\
H&=& \mathcal{G}_{r}^2 -  \mathcal{E}_{r}^2.
\eea

The fourth order polynomial equation (\ref{polinom}) has in general four complex solutions. We give the real four solutions in the permitted range of the circular orbit radii. The representative behavior of the velocity profiles with respect to the radius of the circular orbit for different values of the magnetic field parameter $\BB$ and the rotational parameter $a$ is shown in Fig.\ref{Veloc1}. The real solutions for the magnetized rotating black holes can be related to the solutions corresponding to the non-magnetized black holes, i.e., the circular geodesics. These can correspond to the prograde and retrograde motion that become identical in the special case of the non-rotating black holes. 

\begin{figure*}
\includegraphics[width=\hsize]{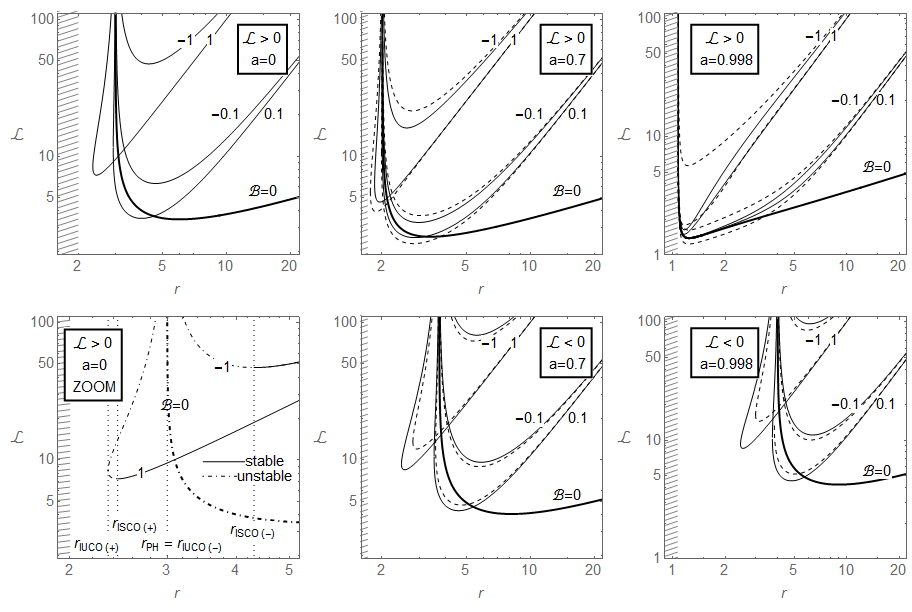}
\caption{\label{AngMomCO}
Radial profiles of the angular momentum of charged particles at the circular orbits, $\LL_{\rm CO}$, given for representative values of the black hole spin $a\in\{0,0.7,0.998\}$ and the magnetic field parameter $\BB\in\{0,\pm0.1,\pm1\}$. The dashed curves correspond to the black hole with zero induced charge $\QQ=0$, while the solid curves represent the case with $\QQ=\QQ_{\rm W}$.
For non-rotating black hole (left lower figure) we give the position of photon orbit and the positions of innermost unstable circular orbit ($r_{\rm IUCO}$ and innermost stable circular orbit $r_{\rm ISCO}$. The $r_{\rm IUCO}$ is boundary of $\LL_{\rm CO}(r)$ function domain and $r_{\rm ISCO}$ is located in $\LL_{\rm CO}(r)$ function minima. 
}
\end{figure*}

For weak magnetic interaction, $\BB<0.1$, the prograde orbits (PALO, PLO) are orbiting in the sense of the black hole rotation, having $v>0$, while the retrograde orbits (RLO, RALO) are orbiting in the inverse sense, having $v<0$; their orientation is opposite relative to the LNRFs. In the case of the Schwarzschild black holes ($a=0$), the prograde and retrograde orbits have the same LNRF-velocity magnitude in both directions at a given radius (the LNRFs correspond to the static frames). In the Kerr spacetimes it is not so, and the range of allowed radii given by the photon circular geodesics, can be extended to smaller distance from the horizon -- see Fig.\ref{Veloc1}.  

Larmor type orbits (with attractive Lorentz force) the velocity sign always corresponds to the $\LL$ sign, while for the anti-Larmor type orbits (with repulsive Lorentz force), the orbits with $\LL>0$ (PALO) can have in some regions $v<0$, if the magnetic interaction parameter is large enough ($\BB>0.1$). The change of LNRF velocity sign will also change the orientation of Lorentz force - if $\BB$ parameter is large enough and velocity become negative $v<0$, then the Lorentz force become attractive even for PALO type of orbits. 
Note that such a change of the orientation of the circular orbits can occur even for the geodesic motion. The family of corotating circular geodesics becomes counterrotating relative to the LNRFs in the ergosphere of Kerr naked singularities with sufficiently small dimensionless spin \cite{Stu:1980:BAC:}.

For a Kerr black hole with given spin $a$, the radial profiles of the charged particle prograde (retrograde) orbits of the Larmor and anti-Larmor type are separated by the radial profile of the prograde (retrograde) circular geodesics, corresponding to the case of $\BB=0$. For the Larmor type orbits, the range of the radii of the prograde (retrograde) charged particle orbits and geodesics is limited from below by the corresponding prograde (retrograde) photon circular geodesic. On the other hand, for the anti-Larmor orbits the radii can enter the region under the related photon circular orbits; if the repulsive Lorentz force is large enough, even stable circular orbits can be located under the radius of the photon circular geodesic, as will be demonstrated below. Therefore, we can introduce the notion of innermost unstable circular orbits (IUCO) located at $r_{\rm IUCO} < r_{\rm ph}$ that give the inner limit on the existence of circular charged particle orbits of the anti-Larmor type with high values of the magnetic parameter $\BB$ -- see Fig.\ref{Veloc1}. For all the four types of the circular orbits (PALO, RLO, PLO, RALO), the LNRF velocity radial profile approaches $v=1$ ($v=-1$, respectively) as the orbits approach the corresponding photon circular geodesic radius representing the inner limit on their existence, or outer limit for orbits with $\BB > 0.1$. 

For both the prograde and retrograde motion, the Larmor and anti-Larmor circular orbits demonstrate qualitatively different radial profiles at large distances from the black hole horizon -- the LNRF velocity of the charged particle orbits becomes ultrarelativistic in the cases of the Larmor circular orbits (PLO and RLO), while it continuously decreases with increasing radius in the case of the anti-Larmor circular orbits (PALO and RALO). Of course, this spectacular effect is caused by the opposite orientation of the Lorentz force in the case of the Larmor and anti-Larmor orbits. The attractive Lorentz force in the Larmor motion supports gravity of the black hole, while the repulsive Lorentz force supports the centrigugal effects in the anti-Larmor motion. 

In the Schwarzschild black hole spacetime the induced charge $\QQ$ vanishes and the regimes PLO with RLO and PALO with RALO become equivalent to each other (see left plot of Fig.\ref{Veloc1}). Moreover, we have to point out that in the case when the black hole spin $a$ and the magnetic interaction parameter $\BB$ are not large, the differences between the two special black hole charge cases, $\QQ=0$ and $\QQ=\QQ_{\rm W}$, are very small, as one can see from Fig.\ref{Veloc1}. However the situation changes dramatically when the black hole spin $a$ is getting close to the extremal value of $a=1$ and the differences of these two cases are profound -- see Fig.\ref{Veloc-grad}. 
%

\subsection{Aschenbach effect around magnetized Kerr black holes}

In the field of non-magnetized ($B=0$) Kerr black holes, the velocity $v(r;a)$ of corotating circular geodesics related to the LNRF (called LNRF velocity) has a monotonous radial profile, if the black hole spin is not too close to the extremal value of $a=1$ -- it increases with decreasing radius, i.e., $\partial v/\partial r < 0$ for all the circular geodesics. However, when the dimensionless spin parameter of the black hole approaches the extremal value, namely when $a>0.9953$, a non-monotonic behaviour has been found in the LNRF velocity profiles of the particles on corotating circular geodesics in the region close to the horizon (see, Fig.\ref{Veloc-grad} for $a=0.998$) \cite{Asch:2004:ASTRA:,Stu-Sla-Tor-Abr:2005:PHYSR4:}. The existence of the LNRF velocity radial profiles changing the sign of velocity gradient and resembling in close vicinity of the black hole horizon a hump has been called ''Aschenbach effect'' \cite{Stu-Sla-Tor:2007:ASTRA:}.

\begin{figure*}
\includegraphics[width=\hsize]{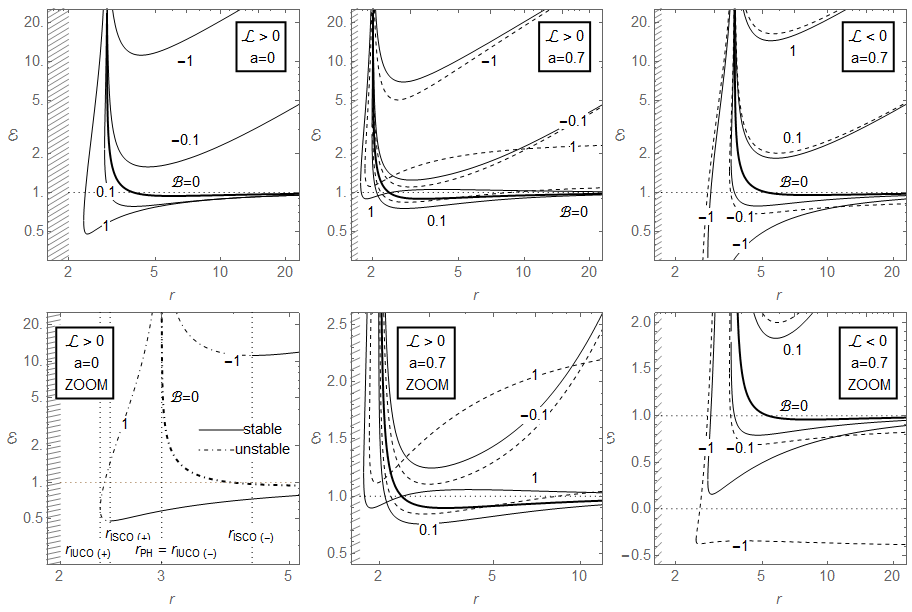}
\caption{\label{EnerCO} Radial profiles of the energy of charged particles at the circular orbits, $\EE_{\rm CO}$, given for representative values of the black hole spin $a\in\{0,0.7\}$ and the magnetic field parameter $\BB\in\{0,\pm0.1,\pm1\}$. The dashed curves correspond to the black hole with zero induced charge $\QQ=0$, while the solid curves represent the case with $\QQ=\QQ_{\rm W}$.
}
\end{figure*}

We have to test existence of the Aschenbach effect in the field of magnetized Kerr black holes, i.e., occurence of the humpy radial profiles of the orbital LNRF velocity of charged particles. However, in the field of magnetized Kerr black holes (or even magnetized Schwarzschild black holes), the orbital LNRF velocity profiles of the Larmor type circular orbits, corresponding to the attractive character of the Lorentz force, have $\partial v/\partial r > 0$ at large distances. There are two types of the Larmor orbits: RLO (with $\LL<0$, $\BB>0$) and PLO (with $\LL>0$, $\BB<0$). 

For both types of the Larmor circular orbits around the magnetized black holes, the radial gradient of the LNRF velocity changes its sign when the contribution of the Lorentz force acting on the charged particle changes from domination to subordination with respect to the effect of the inertial forces due to the angular momentum. Of course, the corresponding local minimum of the LNRF velocity radial profile of the Larmor orbits is not related to the Aschenbach effect, as it occurs even in the field of magnetized Schwarzschild black holes. Moreover, the minimum is located at an intermediate distance from the black hole horizon. On the other hand, it could be relevant to look for possible astrophysical effects related to the general inversion of the LNRF velocity gradient in the family of the Larmor circular orbits. 

For example, we have to study the gradient of the radial profile of the angular frequency of the circular motion, as negative gradient of the radial profile is a crucial condition for the magnetorotational instability \cite{Bal-Haw:1991:ApJ:} governing the viscosity effects in Keplerian discs. Such a condition can be broken in Keplerian discs orbitng in the naked singularity or no-horizon spacetimes \cite{Stu-Sche:2014:CLAQG:,Stu-Sche:2015:IJMPD:}. Surely this condition is also violated at large radii in the Larmor type charged thin discs, as the Larmor orbits demonstrate at large distances the positive gradient of the angular frequency contradicting the MRI condition.

For the circular orbits around magnetized black holes, the Aschenbach effect has to correspond to the existence of a maximum of the LNRF velocity profile, located in vicinity of the black hole horizon. A short analysis has been done for the dependence of the gradient of the orbital velocity $\partial v/\partial r$ on the radius of the circular orbit $r$, in the field of the magnetized black hole with spin $a=0.998$ and the magnetic field parameter $\BB$ -- see Fig.\ref{Veloc-grad}. We demonstrate that the Aschenbach effect really occurs in the field of near-extreme magnetized Kerr blak holes, but only for the prograde Larmor orbits. For the Larmor type orbits the Aschenbach is strenghtened, while for the anti-Larmor type orbits it is suppressed -- see Fig.\ref{Veloc-grad}. No Aschenbach effect has been found for the RLO type orbits. 

\subsection{Energy and angular momentum}

The four solutions for the charged particle circular orbit LNRF velocity $v$ discussed above correspond to the four different types of the circular orbits (PALO, RLO, PLO, RALO). We have thus prograde (retrograde) orbits of the Larmor or Anti-Larmor type. Knowing the LNRF velocity $v$ of these orbits, we are able to determine the specific angular momentum, and the specific energy of these orbits.  

The specific angular momentum of a charged particle following the circular orbit of a given type at a radius $r$ in the field of a magnetized black hole with dimensionless spin $a$, with the electromagnetic interaction characterized by the magnetic parameter $\BB$, can be found from the equation
\beq
\LL_{\rm CO}(r, a, \BB)=\sqrt{g_{\phi\phi}} \gamma\,v\,+\q\,A_{\phi},  \label{Angular}
\eeq
where the Lorentz factor $\gamma=(1-v^2)^{-1/2}$ and the LNRF velocity $v$ is taken as the corresponding solution of Eq.(\ref{polinom}). The four solutions of Eq.(\ref{Angular}) correspond to the Larmor and anti-Larmor motions for the prograde orbits with $\LL>0,v>0$ and retrograde orbits with $\LL<0,v<0$ only for small values of the magntic parameter, $\BB<0.1$, but it is not necessarily so for the anti-Larmor orbits with large values of the magnetic parameter. We give the radial profiles of the specific axial angular momentum of the four types of the charged particle circular orbits (PALO, RLO, PLO, RALO) for typical values of the black hole dimensionless spin $a$ and the magnetic field parameter $\BB$ in Fig.\ref{AngMomCO}. We can convince ourselves that the radial profiles $\LL(r;a,\BB)$ never cross the $\LL=0$ line; therefore, the criterion of the sign of the angular momentum $\LL$ is really convenient for definition of different families of the charged particle circular orbits. 

The specific energy of the charged particle following the circular orbit of the given type can be found analytically from the relation $R= 0$ of (\ref{RfuncDef}) for the given value of the specific angular momenta (\ref{Angular}). On the other hand, the specific energy of the charged particle can be obtained directly from the formalism of forces using the formula \cite{Abr-Nur-Wex:1995:CQG:}
\beq
\EE_{\rm CO}(r, a, \BB) = \sqrt{-g_{tt}} \gamma - \q A_t , \label{enerCO}
\eeq
with the properly chosen value of the LNRF velocity $v$. We give the radial profiles of the specific energy of the four types of the charged particle circular orbits (PALO, RLO, PLO, RALO) for typical values of the black hole dimensionless spin $a$ and the magnetic field parameter $\BB$ in Fig.\ref{EnerCO}. Each point on the radial profiles, giving the circular orbit at a given radius $r$, corresponds to the extremal point of the radial function of the charged particle motion (\ref{RfuncDef}) given for the related specific angular momentum $\LL_{\rm CO}$. Depending on the sign of $\BB$ and $\LL_{\rm CO}$, we can identify the type of the trajectory, namely PALO, RLO, PLO and RALO. In the special case of non-magnetized black hole, $\BB=0$, the motion can be either prograde or retrograde, while in the field of magnetized Schwarzschild black holes, we obtain the Larmor or anti-Larmor motion. We directly see that for the Larmor orbits the specific energy fastly grows with increasing radius, while for the anti-Larmor orbits the specific energy remains at large radii close to the value of $\EE = 1$. For the positive magnetic field parameters, $\BB>0$, the radial profiles of the retrograde (Larmor) orbits are located above the radial profiles of the prograde (anti-Larmor) orbits, while for $\BB<0$, crossing of the radial profiles occurs for black hole spin $a$ large enough, as for the prograde (Larmor) orbits the specific energy strongly increases at large radii -- see Fig.\ref{EnerCO}. We can see that the canonical energy of the circular orbits can be negative, $\EE < 0$, if the magnetic parameter $\BB$ is high enough. 

We notice that the influence of the induced charge on the specific energy of the charged particle at the circular orbit is clearly recognizable -- generally, the induced charge increases the specific energy of the orbit at a fixed radius. Therefore we can conclude that depending on the stage of the accretion of charge into the black hole, the specific energy of the orbiting charged particles $\EE_{\rm CO}$ will be varied. 

\begin{table}
\caption{\label{tabl} Values of the ISCO radius $r_{\rm ISCO}$ in the limiting cases of the non-magnetized, $\BB=0$, and highly magnetized, $\BB\to\infty$, Schwarzschild ($a=0$), Kerr ($a=0.7$) and extreme Kerr ($a=1$) black holes. }
\begin{ruledtabular}
\begin{tabular}{||l||c|cc||c|cc|| }
			 & \multicolumn{3}{c||}{PALO} & \multicolumn{3}{c||}{RALO} \\
	     & $\BB=0$ & \multicolumn{2}{c||}{$\BB\to\infty$} & $\BB=0$ & \multicolumn{2}{c||}{$\BB\to\infty$} \\
			 & 				 & $\QQ_0$ & $\QQ_{\rm W}$ 								&  				& $\QQ_0$ & $\QQ_{\rm W}$ \\
	
 $a=0$	& 6 			& 2 	 & 2 															& 6 	 & 2 		& 2 \\
 $a=0.7$& 3.39 		& 1.72 & 1.78														& 8.14 & 1.72 & 2.24 \\
 $a\to1$& 1 			& 1		 & 1 															& 9		 & 1	  & 2.41 \\
\hline
			 & \multicolumn{3}{c||}{PLO} & \multicolumn{3}{c||}{RLO} \\
	     & $\BB=0$ & \multicolumn{2}{c||}{$\BB\to\infty$} & $\BB=0$ & \multicolumn{2}{c||}{$\BB\to\infty$} \\
			 & 				 & $\QQ_0$ & $\QQ_{\rm W}$ 								&  				& $\QQ_0$ & $\QQ_{\rm W}$ \\
	
 $a=0$	& 6 			& 4.30 & 4.30 													& 6 	 & 4.30 & 4.30 \\
 $a=0.7$& 3.39 		& 2.61 & 2.77 													& 8.14 & 5.60 & 5.45 \\
 $a\to1$& 1 			& 1		 & 1 															& 9 	 & 6.11 & 5.88 \\
\end{tabular}
\end{ruledtabular}
\end{table}

\subsection{Innermost stable circular orbits} \label{Sec-ISCO}

\begin{figure*}
\includegraphics[width=\hsize]{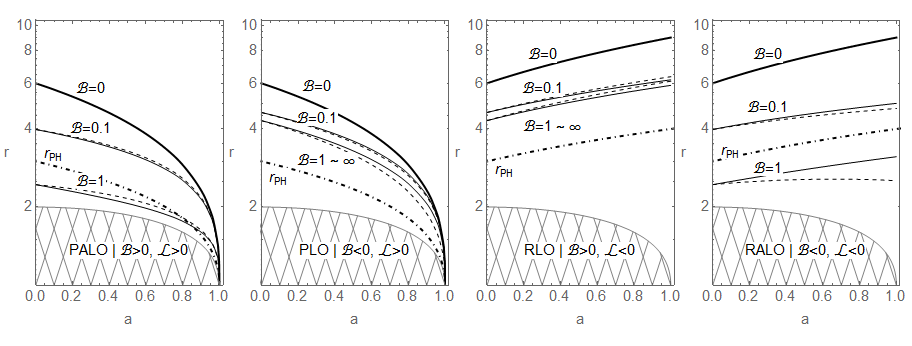}
\caption{\label{RISCO} The ISCO radii given in dependence on the black hole spin $a$ for typical values of the magnetic field parameter $\BB=0,\BB=\pm0.1$ and $\BB=\pm1$. The solid curves correspond to the $\QQ=\QQ_{\rm W}$ case, while the dashed curves are related to the $\QQ=0$ case. The dot dashed curve is representing the radii of photon circular geodesics.
}
\end{figure*}
\begin{figure*}
\includegraphics[width=\hsize]{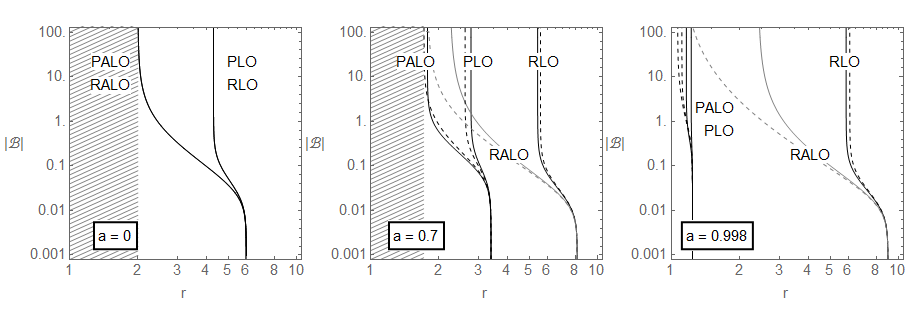}
\caption{\label{RISCOb} 
The ISCO radii given in dependence on the magnetic field parameter $\BB$ for typical values of the black hole spin $a=0, a=0.7$ and $a=0.998$. The curves represent all four classes of the circular orbits. The solid curves correspond to the $\QQ=\QQ_{\rm W}$ case, while the dashed curves are related to the $\QQ=0$ case.
}
\end{figure*}

The local extrema of the radial profiles of the specific angular momentum $\LL_{\rm CO}(r,a,\BB)$, eq. (\ref{Angular}), and the specific energy $\EE_{\rm CO}(r,a,\BB)$, Eq. (\ref{enerCO}), define the innermost stable circular orbits (ISCO). 
Derivative of $\LL_{\rm CO}(r)$ or $\EE_{\rm CO}(r)$ functions with respect to radial coordinate $r$ leads to complicated relations.
It is therefore useful to determine the ISCO position from the radial function equation for radial motion given in the Eq. (\ref{radialmotion}). At the ISCO the the radial function $R(r)$ has to satisfy simultaneously the relations
\beq 
R(r) = 0 \quad \partial_r R = 0, \quad \partial^2_r R = 0. \label{ISCO-cond} 
\eeq
where the function $R(r)$ is given by eq. (\ref{RfuncDef}). We can express (\ref{ISCO-cond}) for any value of black hole charge $\QQ$ in the form
\def\PE{e}
\def\PL{l}
\bea
0 &=& \PE^2 \left(a^2 r+2 a^2+r^3\right)  +4a\PE\PL - \PL^2 \frac{\left(r^2-2r\right)}{r} -\Delta r, \label{eqRR0} \\
0 &=& 4a^3 B\PE  +a^2 [ 4B\PL +\PE (\PE r-2 \QQ)]  \nn \\
	&& +4 B \PL (r-2) r^2 +r^2 \left(3 \PE^2 r-2 \PE \QQ  -2 r+2\right) \nn\\
	&& -r\PL^2 -r\Delta -2a \left(2 B \PE r^2 +\PL \QQ\right), \label{eqRR1} \\
0 &=& 8 -12 r + 8 \BB^2 (6 - 5 r) r^2 + 8 a^2 \BB^2 (3 r-4) \nn\\
	&& +8 \BB \LL (3r-2) + (2 \EE + \QQ) (6\EE r  -4\QQ  +3r\QQ) \nn\\
	&& -8 a \BB (3 \QQ (r-1) + \EE (6r-4)), \label{eqRR2}
\eea
where
\bea
 \PE &=& -\EE + 2 a \BB \,\frac{r-1}{r} + \QQ \,\frac{2-r}{2r}, \\
 \PL &=& \LL -\frac{B}{r} \left(a^2 r-2 a^2+r^3\right) -\frac{a \QQ}{r}.
\eea
The analytic solution to the equations (\ref{eqRR0}-\ref{eqRR2}) can be found for both cases $\QQ=0$ and $\QQ=\QQ_{\rm W}$. For $\QQ=\QQ_{\rm W}$ case one can follow \cite{Ali-Ozd:2002:MNRAS:}, and find angular momentum and the energy $\LL_{\rm ISCO}(r)$ or $\EE_{\rm ISCO}(r)$ as explicit function of $r$ and obtain also one implicit equation for coordinate $r$ - such system must be solved numerically. For $\QQ=0$ case similar solution can be found, having very complicated form. In this article we will obtain the radial position of ISCO by solving the system of equations (\ref{eqRR0}-\ref{eqRR2}) numerically.

The dependence of the ISCO radii on the black hole spin $a$ and the magnetic field parameter $\BB$ is demonstrated in Figs~\ref{RISCO} and \ref{RISCOb} for all the four types of the charged particle circular orbits. 

We can see that in the case of a Kerr black hole with fixed spin $a$, for $\BB>0$ the ISCO radius of the PALO orbits is always lower than the ISCO radius of the RLO orbits, while for $\BB<0$ the ISCO radius of the PLO orbits is lower than the ISCO radius of the RALO orbits for magnitude of $\BB$ low enough, but the PLO ISCO radius becomes larger than RALO ISCO radius for magnitude of $|\BB| > |\BB(a)|$. From the astrophysical point of view it is extremely important that for the anti-Larmor circular orbits, related to the repulsive Lorentz force, the ISCO radius can be lower that the corresponding radius of the photon circular geodesic, $r_{\rm ISCO}<r_{\rm ph}$, if the magnetic parameter $\BB$ is high enough.  

The limiting values of the ISCO radius of the all four classes of the circular orbits, given for limiting values of the black hole spin, $a=0$ and $a=1$, can be found in the Table~\ref{tabl}. Similarly to the case of the innermost stable circular geodesics in the field of extreme Kerr black holes, some of the orbits have the ISCO radius at $r=1$. 

The condition for the particle to be located at the ISCO plays a very important role for the purposes of the present paper. We shall see below that the radial epicyclic frequency vanishes at the ISCO, thus discussion of the oscillatory motion should be restricted to the region limited from below by the ISCO radius. 

\section{Harmonic oscillations of charged particles} \label{Sec-Oscillations}

\begin{figure}
\includegraphics[width=0.9\hsize]{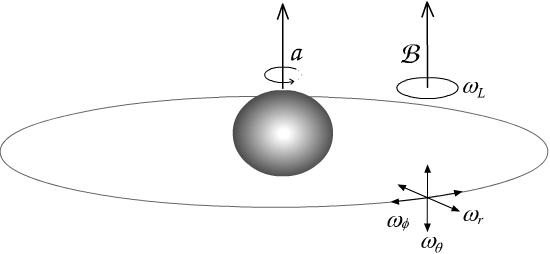}
\caption{ \label{introQPOs}
Locally measured radial (horizontal) $\omega_\mir$, latitudinal (vertical) $\omega_\mit$, Keplerian $\omega_\mip$ and Larmor $\omega_{\mil}$ angular frequencies for charged particle moving in vicinity of a stable circular orbit in the gravitational field of a Kerr black hole combined with an asymptotically uniform magnetic field with strength lines aligned to the rotation axis of the Kerr spacetime.
}
\end{figure}

Stable circular motion of a charged particle revolving at a radius $r_0$ corresponds to a minimum of the radial function $R$ at the radius $r_0$, in the equatorial plane $\theta = \pi/2$, given by Eq.~(\ref{RfuncDef}). If one slightly shifts position of a charged particle from its equilibrium on the circular orbit, the particle starts to oscillate around the position of the radial function minimum, i.e., around the radius of the circular orbit. If the deviation will be small enough, the conditions of the linear harmonic oscillations can be satisfied. The schematic illustration of the frequencies of the charged particle epicyclic oscillations is presented in Fig.~\ref{introQPOs}. We shall study now the small oscillations that can be described as linear harmonic oscillations. 

\begin{figure*}
\includegraphics[width=\hsize]{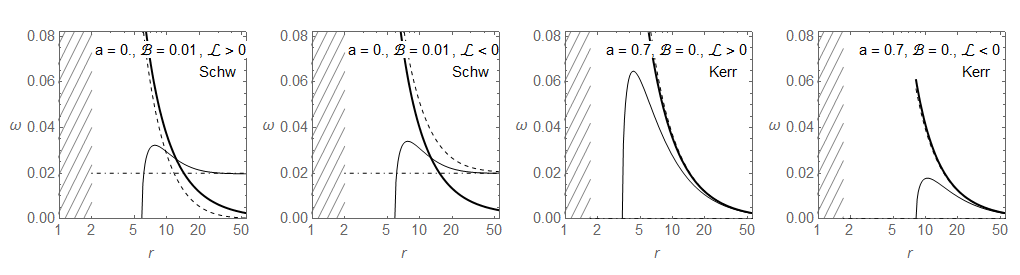}
\includegraphics[width=\hsize]{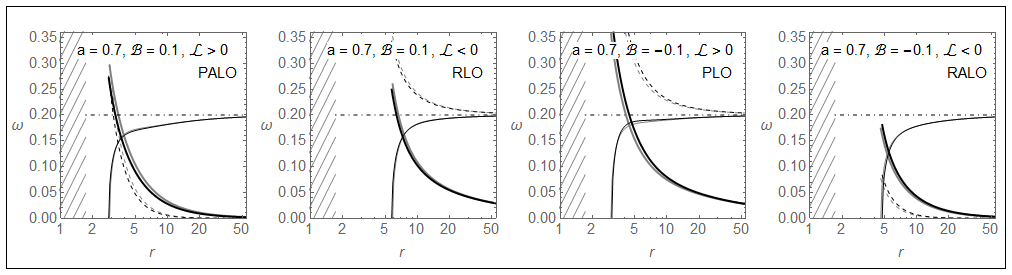}
\includegraphics[width=\hsize]{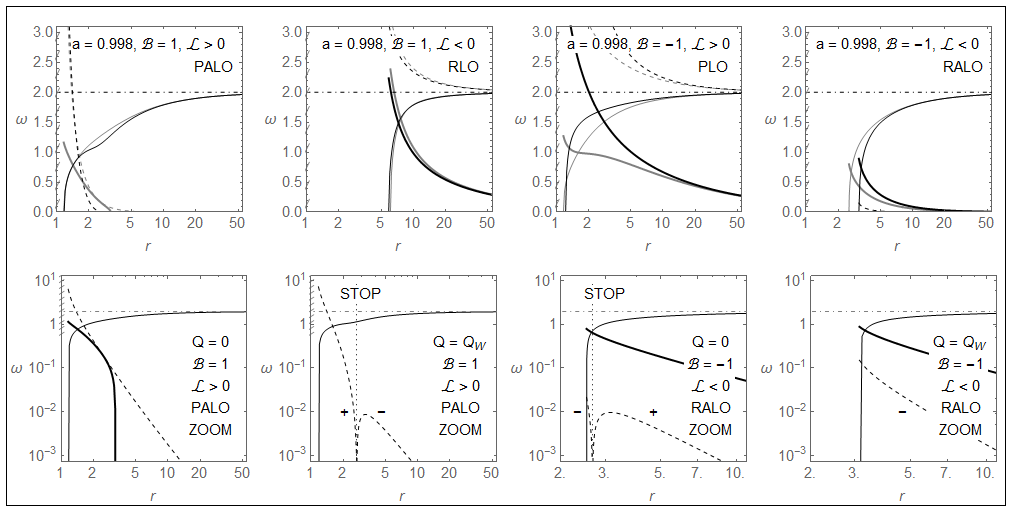}
\caption{\label{ffloc} Radial profiles of the locally measured fundamental frequencies of the charged particle oscillations given for all four classes of the charged particle circular orbits and for typical values of the black hole spin $a$ and the magnetic field parameter $\BB$. 
The curves are drawn in the following way: $\omega_\mir$ -- thin solid, $\omega_\mit$ -- thick solid, $\omega_\mip$ -- dashed, dot-dashed is the Larmour frequency $\omega_{\mil} = 2 \BB$. The black curves denote the case with the induced Wald charge $\QQ=Q_{\rm W}$, while gray curves correspond to zero induced charge, $\QQ=0$. 
The orbital frequency $\omega_\mip=\dot{\phi}$ and can take both positive and negative values according to the sign of angular momentum $\LL$ (we plotted absolute value of $\omega_\mip$). 
%
The first row represent the influence of magnetic field (first two figs) and rotation alone (last two figs). Second row of figures is plotted for middle rotating $a=0.7$ black hole with small value of magnetic field $\BB=\pm0.1$. The transition from $a=0$ or $\BB=0$ cases is smooth and the differences between $\QQ=0$ and $\QQ=\QQ_{\rm W}$ cases are negligible.
Last two rows represents effect of highly rotating $a=0.998$ Kerr black hole in relatively strong magnetic field $\BB=\pm1$. The cases with $\LL\BB<0$ (RLO,PLO) are relatively simple and they are just some smooth modification of figures presented in second row. The difference between $\QQ=0$ and $\QQ=\QQ_{\rm W}$ cases is negligible for RLO, while quite large for PLO. 
Dramatic behaviour of radial profiles can be observed for PALO and RALO cases ($\LL\BB>0$), and we plotted extra figures distinguishing between $\QQ=0$ and $\QQ=\QQ_{\rm W}$ cases for PALO and RALO (last row).
In the PALO case the vertical frequency $\omega_\mir$ is real only for some values of $r$ for $\QQ=0$, while is always complex for for $\QQ=\QQ_{\rm W}$. 
In the PALO $\QQ=\QQ_{\rm W}$ case and in the RALO $\QQ=0$ case the orbital frequency $\omega_\mip$ can change its sign and hence there exist radii where $\omega_\mip$ is zero - at such circular orbit the observer located at infinity will be seeing the charged particle not moving.
}
\end{figure*}

Changes of the position of a charged particle following originally a stable circular orbit can be given in the radial and latitudinal (vertical) directions by variations $\rr = \rr_0 + \dr, \tt = \tt_0 + \dt$ described by the linear harmonic oscillations governed by the equations
\begin{equation}
 \ddot{\dr} + \omega^2_{\mir} ~\dr = 0, \quad \ddot{\dt} + \omega^2_{\mit} ~\dt = 0,
\end{equation}
where dot $\dot{\aaa} = \d \aaa/\d \tau$ denotes the derivative
with respect to 
the proper time $\tau$ of the particle. The locally measured angular frequencies of the radial and latitudinal harmonic oscillatory motion are given by \cite{Wald:1984:book:,Kol-Stu-Tur:2015:CLAQG:}
\beq \label{omrt}
 \omega^2_{\mir}  =  \frac{1}{g_{rr}} \frac{\partial^2 H_{\rm P}}{\partial r^2}, \quad
 \omega^2_{\mit}  =  \frac{1}{g_{\theta\theta}} \, \frac{\partial^2 H_{\rm P}}{\partial \theta^2},
\eeq
where the derivatives are taken from the potential part of the Hamiltonian determined by Eq. (\ref{HamPot}). The energy and angular momentum of the linear harmonic motion are fixed and later equalized to $\EE = \EE_{\rm CO}$ and $\LL = \LL_{\rm CO}$ given by the equations (\ref{Angular}) and (\ref{enerCO}). 
We thus arrive to the relations 
\begin{widetext}
\bea 
\frac{\partial^2 H_{\rm P}}{\partial r^2} \bigg|_{A} &=& \frac{6 a^2 \left(\LL^2-4 \EE^2\right)-8 a \EE \LL (r-4)+4 \LL^2 (r-4)}{2 a^2 \Delta^2}+\frac{4 \left(a^2-1\right) \left[a^2 \left(4 \EE^2-\LL^2\right)+4 a \EE \LL (r-2)-2 \LL^2 (r-2)\right]}{a^2 \Delta^3} \nonumber \\
&&-\frac{2 (\LL-a \EE)^2}{a^2 r^3} +\BB^2 \left(\frac{2 a^2}{r^3}+1\right)-\frac{2 a \BB \QQ}{r^3}+\frac{a^2 \QQ^2}{2 a^2 r^3}, \\
 \frac{\partial^2 H_{\rm P}}{\partial \theta^2}\bigg|_{A} &=& \frac{1}{2 r^3 \Delta}
\Big\{
4 a^5 \BB (\QQ-a\BB) + a^4 \left[4 \EE^2-\QQ^2-2 \BB^2 r (r^2-2r-4)\right] -8 a^3 (\BB \QQ r+\EE \LL) -a^2 \QQ^2 (r^2-2r) \nn\\
&& + 4 a^2 r^2 \left[\EE^2-\BB^2 r(r^2-3r+4)\right] +4 a^2 \LL^2 -4 a \BB \QQ (r-2) r^3+2 (r-2) r^2 \left(\LL^2-\BB^2 r^4\right) 
\Big\},
\eea
\end{widetext}
where $A$ is point in equatorial plane $A=(r,\theta=\pi/2)$.

We have to stress that the radial (horizontal) vanishes at the innermost stable circular orbit, $\omega_{\mir}(r=r_{\rm ISCO})=0$. Further, there exists also the third fundamental angular frequency of the epicyclic particle motion, namely the orbital (axial) angular frequency, $\omega_\mip$, sometimes called the Keplerian frequency, given by the relation 
\beq \label{omp}
 \omega_\mip = \frac{\d \phi}{\d \tau} = \frac{a \left(2 \EE - a \BB r\right) + (r-2) \left(\LL - \BB r^2 \right)}{r \Delta},
\eeq
where ${\d \phi}/{\d \tau} \equiv u^\phi$ is defined by Eq.(\ref{phiequat}) and the specific energy $\EE=\EE_{\rm CO}$ and the specific angular momentum $\LL=\LL_{\rm CO}$ are given by Eqs (\ref{Angular}) and (\ref{enerCO}), respectively. We can notice that the orbital angular frequency does not directly depend on the induced charge $\QQ$, however, the contribution due to the induced charge comes from the dependence of the specific energy $\EE_{\rm CO}$ and the specific angular momentum $\LL_{\rm CO}$ at the circular orbit on the induced charge $\QQ$. 

The pure contribution to the oscillations due to the magnetic field is associated by the Larmor angular frequency $\omega_{\mil}$ which is given by the relation
\beq
\omega_{\mil} = \frac{q B}{m} = 2 |\cb|.
\eeq
Obviously, the Larmor angular frequency $\omega_{\mil}$ does not dependent on the radial coordinate $r$ and it is fully relevant in large distances from the black hole where the uniform magnetic field starts to play the crucial role for the charged particle motion. 

\begin{figure*}
\includegraphics[width=\hsize]{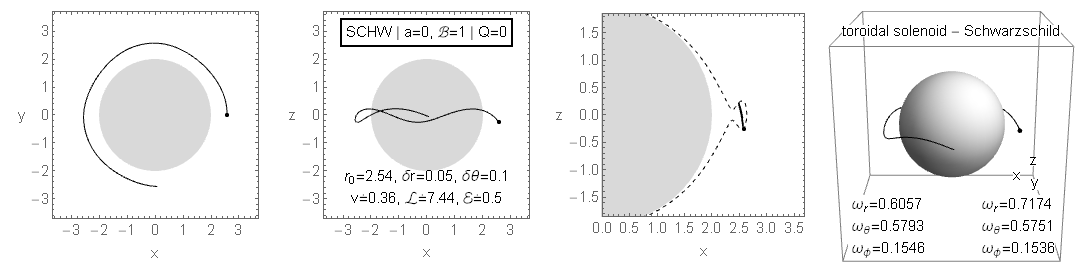}
\includegraphics[width=\hsize]{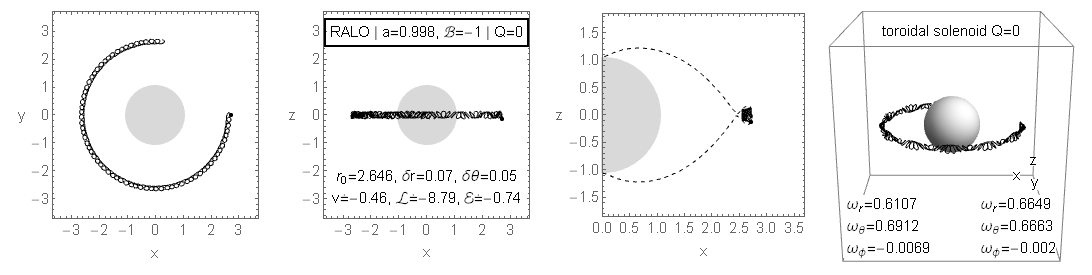}
\includegraphics[width=\hsize]{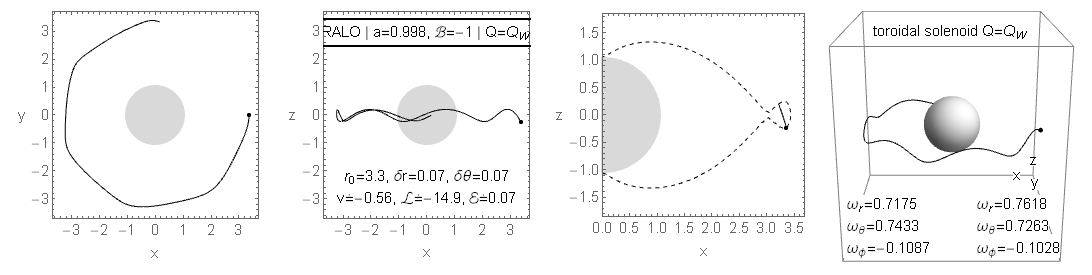}
\caption{\label{traj} 
Charged particles orbits with the toroidal (solenoid) shape, obtained by perturbing the stable circular orbit. 
Toroidal orbits, having almost identical frequencies of the radial and vertical oscillations, $\omega_{\mir} \sim \omega_{\mit}$, and greater then the orbital Keplerian frequency $\omega_{\mip}$, can exist for both non-rotating \Schw{} (first row) and rotating Kerr (second and third row) black holes. 
However, the frequencies $\omega_{\mir} \sim \omega_{\mit} >> \omega_{\mip}$ can occur around magnetized non-rotating \Schw{} black only for large values of the $\BB$ parameter; for rotating Kerr black hole the toroidal orbits are observed in RALO configuration only, see Fig. \ref{ffloc}. Due to vanishing of Keplerian frequency $\omega_{\mip}\sim~0$ close to the $\omega_{\mir}\sim\omega_{\mit}$ radius for $\QQ=0$ RALO configuration, the toroidal shape is more apparent in $\QQ=0$ (second row) then for $\QQ=\QQ_{\rm W}$ (third row). 
In the first column, the polar cap view ($z=0$) on the trajectories is presented. The second column corresponds to the trajectories observed orthogonally to the equatorial plane ($y=0$). The third column shows the boundaries of the particle motion, implied by the condition $H_{\rm P}(r,\theta)=0$, and the cross sections of the orbits observed orthogonally to the equatorial plane ($y=0$). In the fourth column we present the 3D trajectories of the charged particle oscillatory motion. The frequencies in third column are obtained numerically by the Fourier transform of the trajectory (left) and compared with the analytically given frequencies of quasi-harmonic epicyclic motion (right).
}
\end{figure*}
\begin{figure*}
\includegraphics[width=\hsize]{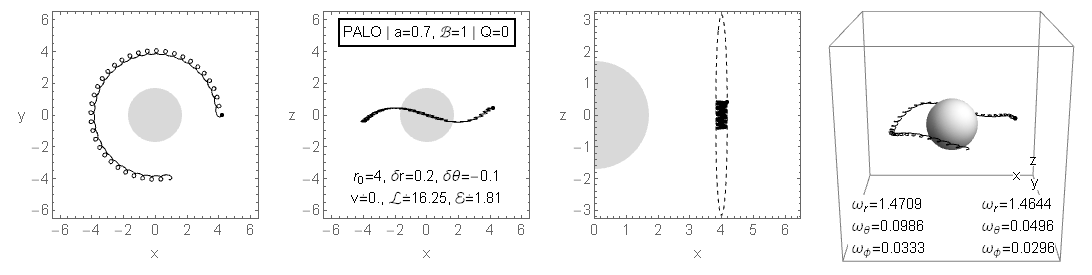}
\includegraphics[width=\hsize]{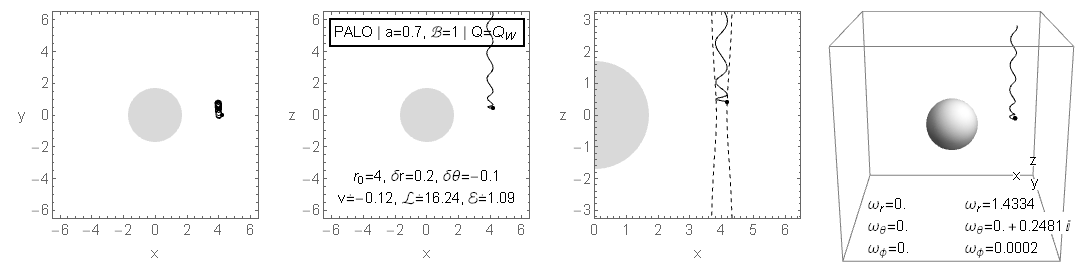}
\caption{\label{lat-inst}
Vertically stable (first row) and unstable (second row) charged particle circular orbits which are stable to radial perturbation (they are located above ISCO). The ISCO for presented PALO configurations with $\LL>0$ and $\BB=1$ is located at $r_{\rm ISCO}\doteq2.6$ for $\QQ=0$ (first row) while at $r_{\rm ISCO}\doteq1.9$ for $\QQ=\QQ_{\rm W}$ (second row). 
The circular orbits with negative LRNF velocity $v < 0$ - see second figure in the first row of Fig. \ref{AngMomCO}, are 
 unstable against vertical perturbations (complex value of $\omega_\mit$) while stable against radial perturbations (real value of $\omega_\mir$). 
%
%
Explanation of presented different views on individual trajectory (diferent subfigures in one row) can be found in Fig. \ref{traj}.
}
\end{figure*}

An alternative definition of the frequencies of the charged particle epicyclic harmonic oscillations is based on variations of the Lorenz equation and can be found in \cite{Ali-Gal:1981:GRG:}. Both the definitions lead to the same results for the radial and latitudinal perturbations, however the frequencies obtained in \cite{Ali-Gal:1981:GRG:} are related to the static distant observer. In order to obtain frequencies measured by distant observer, one needs to divide Eqs (\ref{omrt}) and (\ref{omp}) by the redshift factor $u^{t}$, given by Eq.(\ref{tequat}). We leave the discussion of the frequencies of the charged particle oscillations measured at infinity for the next paper, while in the present paper we concentrate our attention to the properties of the locally measured frequencies, analyzing their radial profiles represented in Fig.~\ref{ffloc}. 

Behavior of the fundamental frequencies $\omega_\mir, \omega_\mit, \omega_\mip, \omega_\mil$ and their ratios can help us to distinguish different shapes of charged particle epicyclic orbits in the vicinity of a stable circular orbit. The representative profiles of the frequencies are given in Fig.~\ref{ffloc}. We compare the frequencies in the cases of magnetized Schwarzschild, Kerr and magnetized Kerr black holes. We can see that there are strong differences between the properties of the oscillatory frequencies around charged particle circular orbits and the geodesic circular orbits. The most relevant difference is related to the fact that the latitudinal frequency, $\omega_\mit$, is not well defined (becomes to be complex) in some regions where the radial motion is stable, if the magnetic parameter $\BB$ is high enough. 
Notice that the charged particle circular orbits are metastable in the flat spacetime with an uniform magnetic field.
Therefore, we demonstrate an instability relative to the vertical perturbations that could enter the problem of the stability of the charged particle circular orbits. Note that this is a new and important phenomenon as the radially stable circular geodesics of the Kerr metric are always stable relative to the vertical perturbations \cite{deF-Cal:1972:NuoCim:,Bic-Stu:1976:BAC:,Stu:1980:BAC:}.

In the case of the charged particle oscillatory motion, the radial, latitudinal and orbital frequencies have to be related to the Larmor frequency. We can summarize their properties in the following way. For oscillations around all four classes of the circular orbits, the radial frequency is always smaller than the Larmor frequency. The orbital frequency is always larger than the Larmor frequency for the Larmor orbits RLO and PLO, while it is smaller than the Larmor frequency at large enough radii for the anti-Larmor orbits PALO and RALO, but it can excedd the Larmor frequency at radii close the horizon radius, if the black hole spin is not close to the extreme value of $a=1$. The latitudinal frequency is much smaller than the Larmor frequency at large radii for all the four types of the circular orbits, and it exceeds the Larmor frequency at radii close to the horizon radius, if the black hole spin is not close to $a=1$. Therefore, the latitudinal frequency radial profile crosses the radial frequency radial profile for all four classes of the circular orbits, while the orbital frequency radial profile crosses the radial frequency radial profile in the case of the anti-Larmor frequencies PALO and RALO. The crossing points are close for the PALO orbits, while they are relatively distant for the RALO orbits. The regions where the latitudinal frequency $\omega_\mit$ is not well defined are located where the velocity of the prograde orbits becomes negative ($\BB > 0, v < 0$). Note that then the instability to the vertical perturbations can imply an escape to infinity, or a bounded chaotic motion. 

The analysis of the oscillatory orbits in the field of magnetized Schwarzschild black holes has been done in \cite{Fro-Sho:2010:PHYSR4:}. Here we extended the study to the case of the magnetized Kerr black holes. For oscillatory motion around circular orbits of all the four classes, the asymptotic values of the frequency of the radial oscillations, $\omega_\mir$, coincides with the Larmor frequency $\omega_\mil$. The same effect occurs for the orbital frequency $\omega_\mip$ in the case of the Larmor orbits, while in the case of the anti-Larmor orbits, $\omega_\mip$ vanishes at infinity. Simultaneously, in the case of the anti-Larmor orbits of both prograde and retrograde type, the influence of the magnetic field can decrease the values of the orbital frequency $\omega_\mip$ down in such a way that they are much less than the corresponding values of the radial and latitudinal frequencies $\omega_\mir$ and $\omega_\mit$. This implies existence of a new type of the trajectories, namely those resembling toroidal (solenoid) orbits with $\omega_r \sim \omega_{\theta} >> \omega_{\phi}$. Such kind of the oscillatory motion is most profoundly demonstrated in the case of the RALO motion. In the Larmor motion of both the PLO and RLO type, no orbits of the toroidal type can be found due to the fact that the frequencies of radial and vertical oscillations become always less than the orbital frequency $\omega_\phi$.

Finally, we give trajectories of the charged particle oscillatory motion near the circular orbits around the magnetized black holes for the qualitatively new type of the toroidal (solenoid) motion allowed by rotating black holes in Fig.~\ref{traj}. The frequencies corresponding to the trajectories represented in Fig.~\ref{traj} are related to the frequencies plotted in Fig.~\ref{ffloc}. The other types of the oscillatory motion can be found in our previous paper \cite{Kol-Stu-Tur:2015:CLAQG:}. 

We further give in Fig.~\ref{lat-inst} trajectories corresponding to the other fundamentally new phenomenon discovered here, namely the vertically unstable motion occuring in the regions of stability against radial perturbations. Such trajectories correspond to perturbed circular orbits of the anti-Larmor type with $\LL>0$ and $v < 0$ and can be both escaping and bounded. 
%

\section{Summary} \label{Summary}

In the present paper we have studied behaviour of the charged test particles in the vicinity of the equatorial plane of a weakly magnetized Kerr black hole. The motion of charged particles, as compared to the geodesic motion, dramatically changes in the presence of even weak magnetic field. 

We demonstrated that the circular motion of charged particles can be separated into four different classes of circular orbits depending on the orientation of the particle motion relative to the black hole rotation, and the orientation of the Lorenz force acting on the charged particles. We presented the qualitative and quantitative analysis of the four classes of the charged particle circular orbits.

We also considered the influence of the induced charge $\QQ$ due to the rotation of the black hole in external magnetic field. We demontrated that this effect is quite small in the case when the spin of the black hole and magnetic field strength are small. However, in the case of fastly rotating black holes and large values of magnetic field, the effect of the induced charge cannot be neglected. 

Using the formalism of forces \cite{Abr-Nur-Wex:1995:CQG:}, we have found the analytical expression for the velocity, specific angular momentum and specific energy of charged particles at the circular orbits. We have shown that far away from the horizon of the black holes, the velocity of charged particles still can be ultrarelativistic, but only in the cases related to the Larmor motion, i.e., for PLO and RLO, while in the anti-Larmor regime (PALO and RALO) the motion can be ultrarelativistic only in the regions close to the black hole horizon. We have also shown that for the prograde motion, the so called Aschenbach effect can be observed in the black hole vicinity. However, there is a change of the velocity gradient also at intermediate distance from the black hole in the case of the Larmor motion. 

We have determined the ISCO orbits for charged particles following all four classes of the circular motion. We have found the numerical values of ISCO radii in the limiting cases, when $\BB >> 1$, and the rotation of the black hole is extremal ($a=1$). We have shown that in the case of prograde motion near rotating black hole, the ISCO radii are always shifted toward the horizon, and in the extremely rotating case its radius coincide with the horizon radius. However, in the retrograde motion there appear situations when the orbits can be shifted outward of the horizon up to $9 M$. 

We have demonstrated that the charged particle circular orbits of the anti-Larmor type, with repulsive Lorentz force, can extend below the radius of the related photon circular geodesic, and such orbits can be even stable against perturbations, if the magnetic field represented by the magnetic parameter $\BB$ is large enough. 

Assuming small deviations of a particle from the equatorial circular orbit and using the method of the perturbation of the Hamiltonian, we studied the harmonic oscillations of charged particles in the uncoupled orthogonal radial and vertical (latitudinal - $\theta$) oscillatory modes. We have found the analytical expressions for the locally measured frequencies of the radial, $\omega_\mir$, vertical, $\omega_\mit$, azimuthal (orbital) $\omega_\mip$ and Larmor, $\omega_\mil$, oscillations. We have studied properties of these frequencies related to the Larmor and anti-Larmor circular orbits of both the prograde and retrograde type. We have found a fundamental new effect related to the instability of the charged particle circular orbits against vertical perturbations

We present the special trajectories of the perturbed circular motion, demonstrating the qualitatively new shape of the charged particle epicyclic motion in the vicinity of stable circular orbits and the types of the vertically perturbed unstable orbits. 

We demonstrate explicitly that in the case of the RALO oscillatory motion a new type of trajectories of the toroidal character (having $\omega_\mir \sim \omega_\mit >> \omega_\mip$) can be obtained if the black hole spin $a$ and the magnetic field parameter $\BB$ have appropriate values. 
The spiral orbits resembling a toroid (solenoid) could generate an internal toroidal magnetic field that could be used as a physical model for axially symmetric current-carrying string loops \cite{Cre-Stu:2013:PHYSRE:}. Such toroid-like orbits have to satisfy the condition $(\omega_\mir\sim\omega_\mit)~\gg~\omega_\mip$, but this condition is not valid for charged particles orbiting a charged source in the weak gravity limit \cite{Kov:2013:EPJP:}, and we have shown that it is not possible to obtain such orbits even for charged particles orbiting a non-rotating \Schw{} black hole placed in an uniform magnetic field. On the other hand, in the field of rotating Kerr black holes and naked singularities, the spiral orbits can exist because of existence of relativistic orbits with low (Keplerian) angular velocity relative to distant static observers \cite{Bal-Bic-Stu:1989:BAC:,Stu:1980:BAC:}.

\section*{Acknowledgments}
The authors acknowledge the Albert Einstein Centre for Gravitation and Astrophysics supported by the Czech Science Foundation grant No. 14-37086G and the Silesian University at Opava grant SGS/23/2013.

\appendix
\section{Electromagnetic fields measured by ZAMO}

One can express the frame components of the external electromagnetic field in the LNRF, i.e., measured by ZAMO as 
\bea
\label{e1} &&E^{\hat{r}}=-\frac{{rBa}\sin \theta }{\Sigma ^{2}A}\left\{ \left[ \Delta
-\left( 1-\frac{M}{r}\right) \Sigma -a^{2}\sin ^{2}\theta \right] \right.  
\nonumber  \\
&&\hspace{1cm}\left. \times (1+\cos ^{2}\theta )\left( r^{2}+a^{2}\right)
+2\Sigma Mr\sin ^{2}\theta \right\} \ , \\
\label{e2} &&E^{\hat{\theta}}=\frac{aB\sin ^{2}\theta }{\Sigma ^{2}\sqrt{\Delta }A}%
\bigg[\{\Delta +2(r^{2}-a^{2})\}(r^{2}+a^{2})\Sigma \cos \theta  
\nonumber \\
&&\hspace{1cm}-\left. \left\{ \Sigma (r^{2}-a^{2})-2aK\cos \theta \right\}
\Delta \right] , \\
\label{m1} &&B^{\hat{r}}=\frac{B\sin 2\theta }{2\Sigma A}\bigg[\Delta a^{2}\sin
^{2}\theta -\frac{2Ka}{\Sigma }(r^{2}+a^{2})  \nonumber \\
&&\hspace{1cm}\left. -(r^{2}+a^{2}\cos 2\theta )(r^{2}-a^{2})\right] , \\
\label{m2} &&B^{\hat{\theta}}=-\frac{rB\sqrt{\Delta }}{\Sigma ^{2}A} 
\left\{ \left[ \Delta -\left( 1-\frac{M}{r}\right) \Sigma -a^{2}\sin ^{2}\theta \right] \right.   \nonumber \\
&&\hspace{1cm}\times a^{2}(1+\cos ^{2}\theta )-\Sigma ^{2}\}\sin ^{2}\theta \ ,
\eea
where
\bea
&& A=\sin \theta \sqrt{(r^{2}+a^{2})^{2}+a^{2}\sin ^{2}\theta }, \\
&& K=\frac{a}{2}\left[\Delta (1+\cos ^{2}\theta )+(r^{2}-a^{2})\sin ^{2}\theta \right].
\eea
In the linear and quadratic approximation in $a$, which has a special interest in the study of the physical phenomena occurring near slowly rotating black holes the expressions (\ref{e1})--(\ref{m2}) take the following form
\bea
&&E^{\hat r} = \frac{B}{r}\left(a\cos^{2}\theta-{M a(1+3\cos 2\theta)}/{2 r}\right) , \\
&&E^{\hat \theta}= \frac{B a \sin\theta}{r}\left(3\cos\theta-1\right) , \\
&&B^{\hat r} =-B\cos\theta\left(1-\frac{a^2}{2r^2} \left(1+3\cos^2\theta\right)\right) , \\
&&B^{\hat\theta} = B\sin\theta\left(1-\frac{M}{r} - \frac{1}{2r^2}(a^2 \cos^2\theta - M^2)\right).  
\eea
The asymptotic values of (\ref{e1})--(\ref{m2}), corresponding to the flat spacetime ($M/r\rightarrow 0$, $Ma/r^2\rightarrow 0$) are simplified to  
\bea
\label{limit_B_1} && \lim_{\rm flat} B^{\hat r}=-B\cos\theta, \quad \lim_{\rm flat} B^{\hat\theta} = B\sin\theta, \nonumber \\
\label{limit_E} && \lim_{\rm flat} E^{\hat r}=\lim_{\rm flat}E^{\hat\theta}=0.
\eea



\def\prc{Phys. Rev. C}
\def\pre{Phys. Rev. E}
\def\prd{Phys. Rev. D}
\def\jcap{Journal of Cosmology and Astroparticle Physics}
\def\apss{Astrophysics and Space Science}
\def\mnras{Monthly Notices of the Royal Astronomical Society}
\def\apj{The Astrophysical Journal}
\def\aap{Astronomy and Astrophysics}
\def\actaa{Acta Astronomica}
\def\pasj{Publications of the Astronomical Society of Japan}
\def\apjl{Astrophysical Journal Letters}
\def\pasa{Publications Astronomical Society of Australia}
\def\nat{Nature}
\def\physrep{Physics Reports}
\def\araa{Annual Review of Astronomy and Astrophysics}
\def\apjs{The Astrophysical Journal Supplement}
\def\aapr{The Astronomy and Astrophysics Review}

\end{document}